\documentclass[aps,prresearch,reprint]{revtex4-2}

\usepackage{amsmath}
\usepackage{amssymb}
\usepackage{graphicx}
\usepackage{physics}
\usepackage{hyperref}
\usepackage{capt-of}

\DeclareMathOperator{\diag}{diag}

\begin{document}

%\preprint{}

\title{Chiral-Mode Control around a Hermitian Diabolic Point \\ in Discrete Non-Hermitian Coupled Resonators}


\author{Kota Yagi\textsuperscript{1,2}, Takahiro Uemura\textsuperscript{1,2}, Yuto Moritake\textsuperscript{1,3}, Adam Mock\textsuperscript{4}, and Masaya Notomi\textsuperscript{1,2,5}}
\affiliation{\textsuperscript{1}Department of Physics, Institute of Science Tokyo, 2-12-1 Ookayama, Meguro-ku, Tokyo 152-8550, Japan}
\affiliation{\textsuperscript{2}NTT Basic Research Laboratories, NTT Inc, 3-1 Morinosato Wakamiya, Atsugi-shi, Kanagawa 243-0198, Japan}
\affiliation{\textsuperscript{3}Institute of Industrial Science, The University of Tokyo, 4-6-1 Komaba, Meguro-ku, Tokyo 153-8505, Japan}
\affiliation{\textsuperscript{4}School of Engineering and Technology, Central Michigan University, Mount Pleasant, Michigan 48859, USA}
\affiliation{\textsuperscript{5}NTT Nanophotonics Center, NTT Inc, 3-1 Morinosato Wakamiya, Atsugi-shi, Kanagawa 243-0198, Japan}


\date{\today}

\begin{abstract} 
Motivated by the prospect of chiral-mode control in compact photonic systems, {{we analyze discrete coupled single-mode resonators.}} Using the minimal three-resonator model, we show that an infinitesimal complex onsite perturbation near a Hermitian diabolic point (DP) induces chiral-mode selection, governed by what we term an asymptotic exceptional point {{(AEP)}}. { Here, an AEP denotes a Hermitian DP equipped with a non-Hermitian perturbation that induces an asymptotically defective effective Hamiltonian. The eigenvectors coalesce in the asymptotic limit toward the DP, although the Hamiltonian at the point itself remains diagonalizable.} {Operationally, this AEP response realizes chirality switching from an achiral state to a chiral state.} The associated eigenvalue response exhibits the anomalous fractional-power scaling $\Delta\lambda \propto \varepsilon^{3/2}$, distinct from the square-root response of {{an ordinary exceptional point (EP)}}.  { We further show that, in a broader two-parameter perturbation space, { ordinary EPs lie on exceptional-line branches that meet at the AEP. A finite-bias control sweep crosses these branches at an EP pair, enabling chirality reversal between opposite chiral states.} The central message is therefore that the AEP organizes two related routes for chirality switching: direct switching from an achiral state to a chiral state via the AEP, and switching between opposite chiral states via an EP pair in the vicinity of the AEP.} Within a finite-resolution averaging model, these two operating points exhibit different practical { performance characteristics}, and under sufficiently high control resolution, the AEP operating point can become more favorable than the EP-pair operating point, suggesting a route toward compact and low-energy chiral photonic devices.
\end{abstract}
%\keywords{}

\maketitle


\section{Introduction}

Physical systems with gain and loss are naturally described by non-Hermitian Hamiltonians {{across a wide range of platforms}}, from optics to acoustics \cite{
zhu-2014-PhysicalReviewX-$mathcalPmathcalT$SymmetricAcoustics, shi-2016-NatureCommunications-AccessingExceptionalPoints, ozdemir-2019-NatureMaterials-ParityTimeSymmetry, el-ganainy-2018-NaturePhysics-NonHermitianPhysicsPT, miri-2019-Science-ExceptionalPointsOptics}. In contrast to Hermitian systems, {{whose eigenvalues and eigenvectors are real and orthogonal, respectively}}, non-Hermitian systems generally exhibit complex eigenvalues and nonorthogonal eigenvectors, giving rise to counterintuitive spectral and dynamical phenomena. Among these, the exceptional point (EP) is a non-Hermitian degeneracy at which both the eigenvalues and the corresponding eigenvectors coalesce. EPs have therefore attracted particular attention as a genuinely non-Hermitian singularity with no Hermitian counterpart \cite{heiss-2012-JournalofPhysicsA:MathematicalandTheoretical-PhysicsExceptionalPoints,
berry-2004-CzechoslovakJournalofPhysics-PhysicsNonhermitianDegeneracies, doppler-2016-Nature-DynamicallyEncirclingExceptional}. 

One topic of growing interest in non-Hermitian physics is the control of chirality \cite{heiss-2001-TheEuropeanPhysicalJournalD-AtomicMolecularOpticalandPlasmaPhysics-ChiralityExceptionalPoints}. Although chirality can be defined in several ways, for example through spin angular momentum, in this work we focus on chirality associated with orbital angular momentum (OAM) \cite{allen-1992-PhysicalReviewA-OrbitalAngularMomentum,yao-2011-AdvancesinOpticsandPhotonics-OrbitalAngularMomentum}. OAM-chiral modes carry clockwise (CW) or counterclockwise (CCW) phase circulation. {{In traveling-wave resonators, such as ring resonators, these circulating modes are commonly referred to as whispering-gallery modes (WGMs) \cite{vahala-2003-Nature-OpticalMicrocavitiesa}.}} They provide a basis for multiplexed optical-vortex communication
% OAM-chiral modes carry clockwise (CW) or counterclockwise (CCW) phase circulation and provide a basis for  multiplexed optical-vortex communication
 \cite{wang-2012-NaturePhotonics-TerabitFreespaceData, willner-2021-AppliedPhysicsReviews-OrbitalAngularMomentum}. {{In rotationally symmetric resonators, CW and CCW modes are degenerate due to time-reversal symmetry, making their selective excitation or isolation a nontrivial task. Recently, non-Hermitian perturbations have emerged as a powerful tool in WGM systems to enable the selective control of these circulating components \cite{zhang-2020-Science-TunableTopologicalCharge,hayenga-2019-ACSPhotonics-DirectGenerationTunable,zhang-2020-Light:Science&Applications-UltrafastControlFractional}}}. Selective control of these modes offers a promising route to manipulating optical flow at the micro- and nanoscale and to applications in photonic and quantum information processing \cite{lodahl-2017-Nature-ChiralQuantumOptics, xu-2023-ScienceAdvances-SubwavelengthControlLight}.

A central mechanism for chiral-mode control in non-Hermitian systems is the EP. At an EP, the effective CW/CCW Hamiltonian becomes defective and cannot be diagonalized, reflecting the emergence of a Jordan-block structure in the Hamiltonian \cite{kato-1995--PerturbationTheoryLinear, heiss-2012-JournalofPhysicsA:MathematicalandTheoretical-PhysicsExceptionalPoints}. 
{{In waveguides with complex periodic perturbations formed by coupled index and gain–loss gratings, asymmetric propagation near an EP originates from directional asymmetry in Bragg backscattering, producing one-sided reflection suppression \cite{lin-2011-PhysicalReviewLetters-UnidirectionalInvisibilityInduced, yin-2013-NatureMaterials-UnidirectionalLightPropagation}. { The same directional asymmetry mechanism for traveling waves} carries over to microring resonators, where the forward/backward scattering channels are mapped onto the CW and CCW cavity modes, enabling chiral mode formation near the corresponding EP \cite{miao-2016-Science-OrbitalAngularMomentum}.}}
{{In addition, chiral-mode lasing and switchable directional emission have also been demonstrated through EP-induced asymmetric scattering engineered by controlling nanoscale scatterers \cite{wiersig-2011-PhysicalReviewA-StructureWhisperinggalleryModes, peng-2016-ProceedingsoftheNationalAcademyofSciences-ChiralModesDirectional,svela-2020-Light:Science&Applications-CoherentSuppressionBackscattering,lee-2023-eLight-ChiralExceptionalPoint}.}}
% These demonstrations mainly rely on comparatively large WGM resonators, so further miniaturization is desirable for on-chip photonics\cite{vahala-2003-Nature-OpticalMicrocavitiesa,peng-2016-ProceedingsoftheNationalAcademyofSciences-ChiralModesDirectional}.

In parallel with these developments, the exploration of chiral modes has begun to extend beyond large-footprint WGM resonators toward more compact and lumped systems \cite{mock-2019-ACSPhotonics-MagnetFreeCirculatorBased, fong-2021-PhysicalReviewResearch-ChiralModesExceptional}. We also found that chiral modes can emerge abruptly from originally achiral modes in waveguide structures when {complex perturbations, involving both refractive-index and loss modulations,} are introduced perturbatively\cite{Takiguchi_CLEO2026}. {{While chiral-mode control has been extensively investigated in traveling-wave WGM systems \cite{peng-2016-ProceedingsoftheNationalAcademyofSciences-ChiralModesDirectional, sweeney-2019-PhysicalReviewLetters-PerfectlyAbsorbingExceptional, chen-2017-Nature-ExceptionalPointsEnhance, zhu-2023-PhysicalReviewA-LocalChiralityExceptional}, a systematic understanding of such control in discrete, coupled single-mode resonators remains elusive. {These coupled-resonator systems, composed of discrete controllable elements and interactions, are ideal for clarifying the fundamental physics of chirality emergence. They are also well suited for the ultimate miniaturization of chiral photonic devices and provide a versatile route toward more complex large-scale systems} \cite{lodahl-2017-Nature-ChiralQuantumOptics, xu-2023-ScienceAdvances-SubwavelengthControlLight}.}}

Motivated by this {{perspective}}, we study { discrete coupled single-mode resonators as a minimal localized-mode platform} for { chirality switching of OAM modes}. We begin with { a} three-resonator system, which is the smallest discrete model supporting a degenerate pair of opposite-circulation modes. { Throughout this work, we quantify the OAM chirality of a localized eigenmode by the imbalance between its projections onto the two opposite-circulation components of this degenerate pair.}  { In this setting, each resonator is represented by a lumped single-mode degree of freedom, allowing chirality emergence to be analyzed within a minimal finite-dimensional model.}

Using { temporal coupled-mode theory} \cite{h.a.haus-1991-ProceedingsoftheIEEE-CoupledmodeTheory,yariv-1999-OpticsLetters-CoupledresonatorOpticalWaveguidea}, {{we show that an infinitesimal complex onsite perturbation near a Hermitian diabolic point (DP) induces chiral-mode selection governed by what we term an asymptotic exceptional point (AEP). { In this work, an AEP denotes a Hermitian DP equipped with a non-Hermitian perturbation structure { that induces an asymptotically defective effective Hamiltonian. Under this perturbation}, the eigenvectors acquire EP-like coalescence asymptotically as the perturbation strength tends to zero.} { In operational terms, this mechanism realizes chirality switching from an achiral state to a chiral state.} The associated eigenvalue response exhibits an anomalous fractional-power scaling $\Delta\lambda \propto \varepsilon^{3/2}$, distinct from the square-root response $\Delta \lambda \propto \varepsilon^{1/2}$ of an ordinary EP \cite{kato-1995--PerturbationTheoryLinear,heiss-2012-JournalofPhysicsA:MathematicalandTheoretical-PhysicsExceptionalPoints} realized at a finite parameter value.}} { We then extend the perturbation setting around { this AEP} and show that the degenerate chiral-mode pair also supports chirality reversal through ordinary EPs. { In the two-parameter perturbation space, the ordinary EPs lie on exceptional-line branches that meet at the AEP.} { The central result is therefore that the AEP organizes two related principles of chirality switching: chiral-mode selection from an achiral state via the direct AEP response, and chirality reversal between opposite chiral states via an ordinary EP pair in the vicinity of the AEP.}}

{The remainder of this paper is organized as follows. Section II introduces the minimal three-resonator model and } { establishes the first switching principle, switching from an achiral state to a chiral state via an AEP}. { 
Section III then considers an extended perturbation setting and { establishes the second switching principle, switching between opposite chiral states via { an EP pair in the vicinity of the AEP.}} Section IV discusses these two operating principles from the viewpoint of chiral-mode control, compares their operating points under finite-resolution operation, and summarizes their practical implications and limitations. Section V concludes, and the Appendices provide the effective-Hamiltonian derivations, the N-site generalization, and the finite-resolution analysis.}

\section{Minimal three-resonator model and chirality generation via an AEP}
\label{sec2:chiral_mode_emergence}

\subsection{Hermitian three-resonator model and degenerate subspace}

\begin{figure}[tpb]
  \centering
  \includegraphics[width=\linewidth]{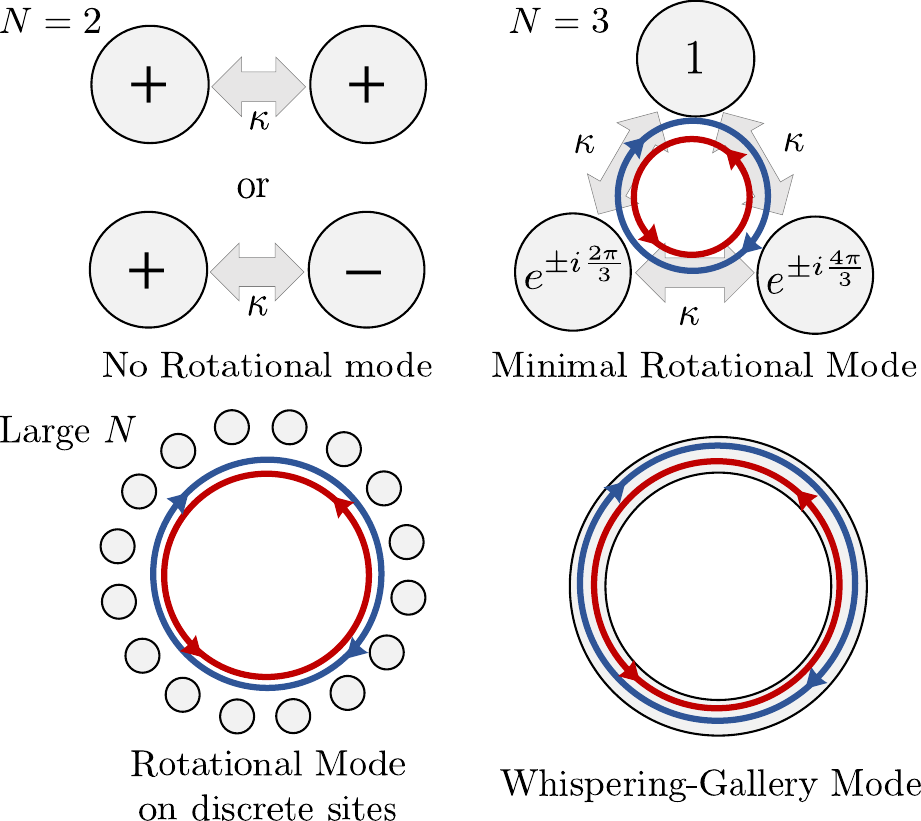}
  \caption{Schematic of coupled resonator systems and mode chirality. In a two-resonator system, the fundamental modes consist of a symmetric mode (in-phase) and an anti-symmetric mode (phase difference of $\pi$), neither of which possesses chirality. The minimal configuration for supporting chiral modes is a three-resonator system. For coupled systems where $N \geq 3$, discrete resonators exhibit rotational modes that correspond to clockwise (CW) and counter-clockwise (CCW) whispering-gallery modes.}
  \label{fig:chiralmode}
\end{figure}

{{To clarify the first AEP-based principle of chirality switching considered in this work,}} we begin with the minimal discrete model that supports chiral eigenmodes: a ring of three coupled single-mode resonators as shown in Fig.~\ref{fig:chiralmode}. In a two-resonator system, the eigenmodes consist only of an in-phase symmetric mode and an out-of-phase antisymmetric mode, neither of which represents a rotating state. 
By contrast, for $N \ge 3$, the discrete rotational symmetry gives rise to a pair of degenerate eigenmodes associated with opposite circulation.
{{The three-resonator system is therefore the minimal discrete platform in which OAM chiral modes can be defined and analyzed analytically.}}
These rotating modes are discrete counterparts of the CW and CCW modes of conventional WGM resonators, in the sense that their phases wind across the resonators. In the case of $N=3$, the corresponding rotating modes are given by phase increments of $e^{i2\pi/3}$ around the three resonators. 

Within the framework of { temporal coupled-mode theory} \cite{h.a.haus-1991-ProceedingsoftheIEEE-CoupledmodeTheory,yariv-1999-OpticsLetters-CoupledresonatorOpticalWaveguidea}, we denote the complex field amplitudes of the three resonators by
\begin{align}
  \ket{a(t)} = (a_1(t),a_2(t),a_3(t))^\mathsf{T},
\end{align}
where all resonators are assumed to share the same resonance frequency $\omega_0$, the coupling rate between adjacent resonators is denoted by $\kappa$, and the superscript $\mathsf{T}$ denotes matrix transposition.

The dynamics is then governed by
\begin{align}
  i \dv{t} \ket{a(t)}
  =
  \mqty(
  \omega_0 & \kappa   & \kappa   \\
  \kappa   & \omega_0 & \kappa   \\
  \kappa   & \kappa   & \omega_0
  )
  \ket{a(t)},
  \label{eq:3mode_dynamics}
\end{align}
where we denote the Hamiltonian on the right-hand side by $H_0$. Its eigenvalues and normalized eigenvectors are
\begin{align}
  \begin{alignedat}{2}
    \lambda_1 & = \omega_0 + 2\kappa, \qquad &
    \ket{L_0} & = 
    \mqty(
    1                                          &
    1                                          &
    1
    )^\mathsf{T}/\sqrt{3},                                         \\
    \lambda_2 & = \omega_0 - \kappa, \qquad  &
    \ket{L_+} & = 
    \mqty(
    1                                          &
    e^{i\frac{2\pi}{3}}                        &
    e^{i\frac{4\pi}{3}}
    )^\mathsf{T}/\sqrt{3},                                         \\
    \lambda_3 & = \omega_0 - \kappa, \qquad  &
    \ket{L_-} & = 
    \mqty(
    1                                          &
    e^{-i\frac{2\pi}{3}}                       &
    e^{-i\frac{4\pi}{3}}
    )^\mathsf{T}/\sqrt{3}.
  \end{alignedat}
  \label{eq:3mode_eigenvalues_local}
\end{align}
Since $\ket{L_+}$ and $\ket{L_-}$ are degenerate, the basis within this two-dimensional eigenspace is not unique. The point $\lambda=\lambda_\pm=\omega_0-\kappa$ is therefore a Hermitian degeneracy, corresponding to a diabolic point (DP). { Hereafter, we use the term chirality specifically in the sense of OAM chirality for localized modes: whether a perturbed eigenmode is biased toward one or the other member of the degenerate opposite-circulation pair, $\ket{L_+}$ or $\ket{L_-}$. In this discrete resonator setting, these two states play the role of the CW/CCW OAM doublet familiar from traveling-wave systems, but are realized here as localized coupled-resonator modes with opposite phase winding. The problem addressed in this work is therefore to determine how non-Hermitian perturbations control these two components of the degenerate OAM pair.} { At the DP, the two-dimensional degenerate subspace does not define a unique handedness for an individual eigenstate; in an operational sense, the unperturbed state is treated here as an achiral starting point for chirality switching.}

Throughout this work, we measure frequencies in units of $\kappa$ and use dimensionless Hamiltonians and eigenvalues. Specifically, under the rescalings $(H_0-\omega_0)/\kappa \to H_0$ and $(\lambda_j-\omega_0)/\kappa \to \lambda_j$, we obtain
\begin{gather}
  H_0=
  \mqty(
  0 & 1 & 1 \\
  1 & 0 & 1 \\
  1 & 1 & 0
  ), \\
  \lambda_1=2,
  \qquad
  \lambda_2=-1,
  \qquad
  \lambda_3=-1.
\end{gather}
Unless otherwise stated, we use this dimensionless representation hereafter.

\begin{figure*}[tbph]
  \centering
  \includegraphics[width=\linewidth]{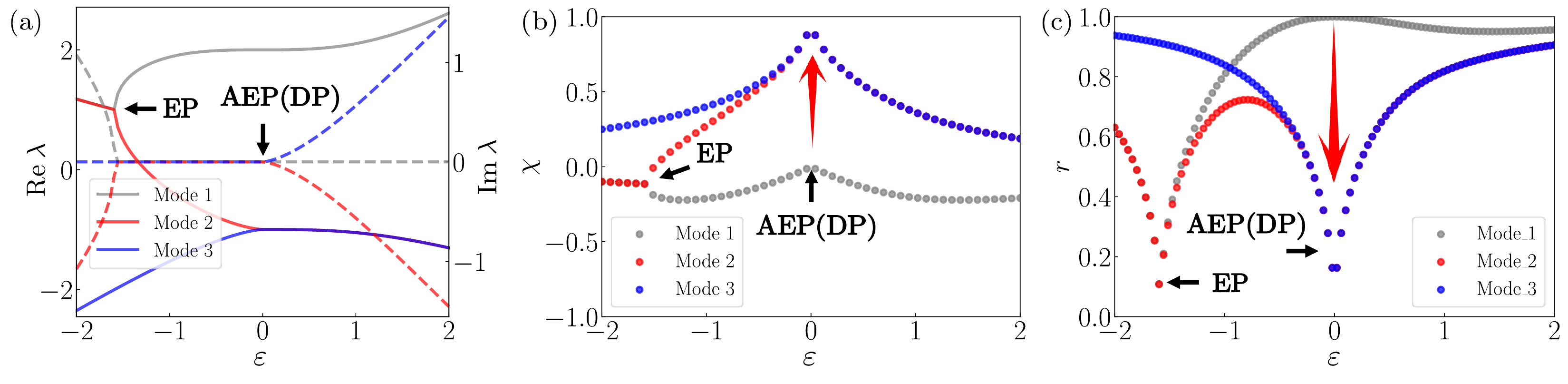}
  \caption{Results for $H=H_0+\varepsilon V_+$ { with real $\varepsilon$. The label AEP (DP) denotes the Hermitian { achiral point}.}  (a) Real parts (solid curves) and imaginary parts (dashed curves) of the eigenvalues $\lambda_j$ as functions of $\varepsilon${{, showing both the DP at $\varepsilon=0$ and the EP at $\varepsilon=-\sqrt[3]{4}$}}. (b) Chirality $\chi_j$ as a function of $\varepsilon$. Although the eigenvalues vary continuously, the eigenstates exhibit singular chiral-mode selection in the limit $\varepsilon\to0$. (c) Phase rigidities as functions of $\varepsilon$, showing that the singularity becomes increasingly pronounced as $\varepsilon\to0$.}
  \label{fig:V+}
\end{figure*}

\subsection{Complex phase-circulating onsite perturbation}

{ We now introduce a non-Hermitian perturbation to ask whether the DP state, whose chirality is not uniquely defined, can be driven into a chiral state by an infinitesimal complex onsite modulation. In other words, our goal in this section is to establish switching from an achiral state to a chiral state {through the response analyzed below.}} Motivated by gain--loss grating structures used to control chiral modes in previous studies \cite{miao-2016-Science-OrbitalAngularMomentum},
{ as well as phase-gradient modulation strategies in discrete resonator arrays, we seek to extend these concepts to our three-resonator system. In particular, drawing inspiration from $2\pi/3$ phase-staggered spatiotemporal modulations \cite{mock-2019-ACSPhotonics-MagnetFreeCirculatorBased} which demonstrate that a cyclic phase pattern across resonators can endow the degenerate subspace with opposite handedness, we capture this mechanism here in a static non-Hermitian form. Specifically, we introduce the complex onsite perturbation:}
\begin{align}
  H = H_0 + V = H_0 + \varepsilon V_\pm,
  \label{eq:3mode_perturbation}
\end{align}
where $\varepsilon$ is the perturbation strength normalized by the coupling {{rate}}. The perturbation matrices $V_+$ and $V_-$ are defined as
\begin{align}
  V_\pm & =
  \mqty(
  1   & 0            & 0            \\
  0   & e^{\mp i2\pi/3}  & 0            \\
  0   & 0            & e^{\mp i4\pi/3}
  ).
  \label{eq:3mode_perturbation_H_pm}
\end{align}
The real parts of the diagonal elements represent resonance-frequency modulation at each resonator, while the imaginary parts represent gain and loss. The perturbation therefore realizes a site-dependent complex onsite modulation with cyclic phase winding. Importantly, the perturbation acts only on onsite terms and does not require nonreciprocal couplings.

For clarity, we work in the eigenbasis of Eq.~\eqref{eq:3mode_eigenvalues_local}. We denote their representation in the site basis by $[A]_{\mathrm{site}}$ and $[\ket{\psi}]_{\mathrm{site}}$ for an arbitrary matrix $A$ and a state vector $\ket{\psi}$. Let
\begin{align}
  U=\mqty(\ket{L_0} & \ket{L_+} & \ket{L_-}),
\end{align}
and define the transformed state and Hamiltonian as
\begin{align}
  \begin{aligned}
      [\ket{a(t)}]_L = U^\dagger [\ket{a(t)}]_{\mathrm{site}},
  &&
  [H]_L = U^\dagger [H]_{\mathrm{site}} U,
  \end{aligned}
\end{align}
where $\dagger$ denotes the Hermitian conjugate. Then the basis vectors and the Hamiltonian in the eigenbasis take the form
\begin{gather}
  \begin{aligned}
    \relax[\ket{L_0}]_L=
    \mqty(
    1 \\ 0 \\ 0
    ),
    &&
    [\ket{L_+}]_L=
    \mqty(
    0 \\ 1 \\ 0
    ),
    &&
    [\ket{L_-}]_L=
    \mqty(
    0 \\ 0 \\ 1
    ), \end{aligned} \notag \\
    \relax[H_0]_L=
    \mqty(
    2 & 0 & 0 \\
    0 & -1 & 0 \\
    0 & 0 & -1
    ),  \label{eq:3mode_dynamics_mock} \\
    \begin{aligned}
          \relax[V_+]_L=
    \mqty(
    0 & 1 & 0 \\
    0 & 0 & 1 \\
    1 & 0 & 0
    ),
    &&
    \relax[V_-]_L=
     \mqty(
    0 & 0 & 1 \\
    1 & 0 & 0 \\
    0 & 1 & 0
    ).
  \end{aligned}\notag
\end{gather}

{ To quantify the OAM chirality of the right eigenstates $\ket{a_j}$ of the Hamiltonian $H$, we project each perturbed eigenstate onto the unperturbed degenerate OAM basis $\{\ket{L_+},\ket{L_-}\}$ and define}
\begin{align}
  \chi_j
  :=
  \qty|\braket{L_+}{a_j}| - \qty|\braket{L_-}{a_j}|,
  \qquad j=1,2,3.
  \label{eq:chirality_definition}
\end{align}
{ This quantity measures the imbalance between the two opposite phase-winding components of the degenerate OAM doublet. The right eigenvectors are normalized as $\braket{a_j}{a_j} = 1$ before evaluating $\chi_j$. Thus, $\chi_j=+1$ ($-1$) corresponds to a mode purely aligned with $\ket{L_+}$ ($\ket{L_-}$), while $\chi_j=0$ indicates equal participation of the two opposite-circulation components and hence no net OAM chirality. In this sense, $\chi_j$ is the localized-mode analogue of a CW/CCW modal imbalance, specialized here to the degenerate OAM pair of the coupled-resonator system.}

Fig.~\ref{fig:V+}(a)--(c) shows the eigenvalues $\lambda_j$, the chirality $\chi_j$, and the phase rigidity $r_j:= {|\braket{\tilde{a}_j}{a_j}|}/{\sqrt{\braket{\tilde{a}_j}{\tilde{a}_j}\braket{a_j}{a_j}}}$,
% \begin{align}
%   r_j
%   :=
%   \frac{|\braket{\tilde{a}_j}{a_j}|}
%   {\sqrt{\braket{\tilde{a}_j}{\tilde{a}_j}\braket{a_j}{a_j}}},
% \end{align}
as functions of $\varepsilon$. Here, $\ket{a_j}$ and $\bra{\tilde a_j}$ denote the right and left eigenstates, respectively. 
The phase rigidity quantifies the nonorthogonality of left and right eigenstates in a non-Hermitian system. {{In non-Hermitian optics, the closely related Petermann factor $K_j$ is often used as a measure of the mode nonorthogonality \cite{berry-2003-JournalofModernOptics-ModeDegeneraciesPetermann}; with the present definition, they are related by $K_j = r_j^{-2}$.}} In a Hermitian system, left and right eigenstates are Hermitian conjugates of each other, {{so that $r_j = 1$ (equivalently, $K_j = 1$)}}. In contrast, non-Hermitian systems generally satisfy $0\le r_j<1$. At an EP, one has self-orthogonality, $\braket{\tilde a_j}{a_j}=0$, and therefore $r_j=0${{; equivalently, the Petermann factor diverges.}} The decrease of $r_j$ thus serves as a standard indicator that the system is approaching an EP \cite{bulgakov-2006-PhysicalReviewE-PhaseRigidityAvoided, wiersig-2023-PhysicalReviewResearch-PetermannFactorsPhase,schomerus-2024-PhysicalReviewResearch-EigenvalueSensitivityEigenstate}. As shown in Fig.~\ref{fig:V+}(a), the eigenvalues along the $\varepsilon$-axis feature both a Hermitian degeneracy at the DP ($\varepsilon=0$) and an EP at a finite negative value of $\varepsilon = -\sqrt[3]{4}$. This EP corresponds primarily to the coalescence of modes originating from $\ket{L_0}$; consequently, the resulting eigenmode does not exhibit OAM chirality.

The most striking feature appears in the eigenstates near the DP. Fig.~\ref{fig:V+}(b) shows that, although the unperturbed system at $\varepsilon=0$ does not select either chirality, an arbitrarily small nonzero perturbation immediately generates an eigenstate with nearly maximal chirality, $\chi\simeq 1$. { Operationally, this corresponds to switching from an achiral state to a chiral state enabled by chiral-mode selection within the degenerate subspace.} Notably, while the eigenvalues evolve continuously from the DP, the eigenvectors undergo an abrupt selection in the limit $\varepsilon \to 0$. As shown in Fig.~\ref{fig:V+}(c), this discontinuity is fundamentally linked to the vanishing phase rigidity, reflecting an EP-like self-orthogonality.

These observations imply that the DP under the complex perturbation $V_+$ acquires an asymptotically EP-like character: the pair of states associated with $\ket{L_+}$ and $\ket{L_-}$ effectively coalesces into a single chiral mode as $\varepsilon\to0$. In other words, the perturbation $V_\pm$ triggers an abrupt coalescence of the eigenstates
\begin{align}
  \qty{\ket{a_1},\ket{a_2},\ket{a_3}}
  \to
  \qty{\ket{L_0},\ket{L_\pm},\ket{L_\pm}},
  \label{eq:3mode_coalesce_show}
\end{align}
for any infinitesimal $\varepsilon\neq0$. At exactly $\varepsilon=0$, the two-dimensional degenerate eigenspace at $\lambda=-1$ can be represented by arbitrary orthogonal linear combinations of $\ket{L_+}$ and $\ket{L_-}$. {{Consequently, the chirality of individual eigenstates is not uniquely defined. However, a finite non-Hermitian perturbation resolves this arbitrariness by selecting an asymptotically chiral direction in the degenerate subspace.}} { Hereafter, we call this Hermitian DP, together with the non-Hermitian perturbation structure that produces the asymptotic coalescence, an asymptotic exceptional point (AEP). The AEP itself remains a diagonalizable Hermitian DP at $\varepsilon=0$, but its eigenvectors exhibit EP-like coalescence in the limit of infinitesimally small but nonzero $\varepsilon$.}

\begin{figure}[t]
  \centering
  \includegraphics[height = 4.8cm]{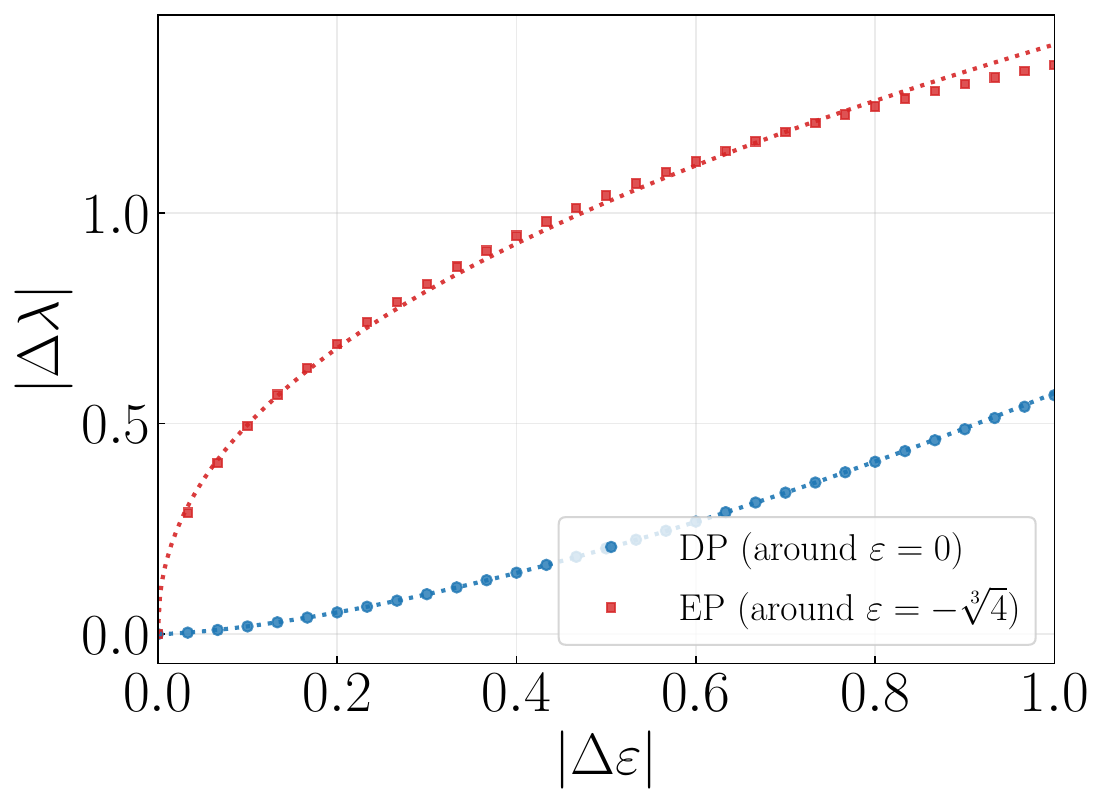}
  \caption{Scaling of the eigenvalue splitting $|\Delta\lambda|$ as a function of the perturbation strength $|\Delta\varepsilon|$ near the DP and the EP. The numerical data (symbols) are in agreement with the predicted scaling laws $|\Delta\lambda| \propto |\Delta\varepsilon|^{3/2}$ for the DP and $|\Delta\lambda| \propto |\Delta\varepsilon|^{1/2}$ for the EP. Dashed lines represent the numerical fits: $|\Delta\lambda_\mathrm{DP}| = 0.5723 \times |\Delta \varepsilon|^{1.495}$ and $|\Delta\lambda_\mathrm{EP}| = 1.401 \times |\Delta \varepsilon|^{0.4496}$.}
  \label{fig:lambda_fit}
\end{figure}

{{Furthermore, the response near the DP is qualitatively different from that near an ordinary EP: instead of the square-root law $|\lambda-\lambda_{\mathrm{EP}}|\propto \Delta \varepsilon^{1/2}$, the present perturbation instead yields the anomalous scaling $|\lambda-\lambda_{\mathrm{DP}}|\propto \Delta \varepsilon^{3/2}$, as shown in Fig.~\ref{fig:lambda_fit}. This behavior is distinct from both the square-root EP response and the linear lifting of a Hermitian degeneracy under a Hermitian perturbation. This anomalous scaling and the associated chiral-mode selection will be derived analytically in the next subsection.}}

\subsection{Analytical derivation and physical origin of the \texorpdfstring{$\varepsilon^{3/2}$}{ε³ᐟ²} law}

{{To clarify the physical mechanism underlying the AEP, we now examine the system analytically.}}
The perturbative eigenvalue response can be obtained analytically from the characteristic equation
{{\begin{align}
  \det(H-\lambda I)=(\lambda+1)^2(\lambda-2)-\varepsilon^3=0.
\end{align}
This characteristic equation is identical for both cases, $V=\varepsilon V_+$ and $V=\varepsilon V_-$.}}
For the two modes associated with the degenerate subspace, {{solving this equation perturbatively near $\lambda=-1$,}} one finds
\begin{align}
  \lambda_{2,3}(\varepsilon)
  =
  -1 \pm \frac{i}{\sqrt{3}}\varepsilon^{3/2} + O(\varepsilon^3).
  \label{eq:3mode_eigenvalue_perturbation}
\end{align}
Equation~\eqref{eq:3mode_eigenvalue_perturbation} confirms that the perturbation gives rise neither to the square-root splitting of EPs nor to the linear lifting expected for an ordinary Hermitian DP, but instead to the anomalous fractional-power scaling $\varepsilon^{3/2}$.

For $V=\varepsilon V_+$, the corresponding eigenvectors behave as
\begin{align}
  \begin{aligned}
    \relax[\bra{\tilde a_{2,3}(\varepsilon)}]_L
                                            & \sim
    \mqty(
    -\frac{1}{3}\varepsilon                 &
    \pm \frac{i}{\sqrt{3}}\varepsilon^{1/2} &
    1
    ),
    \\
    [\ket{a_{2,3}(\varepsilon)}]_L
                                            & \sim
    \mqty(
    -\frac{1}{3}\varepsilon                        &
    1                                              &
    \pm \frac{i}{\sqrt{3}}\varepsilon^{1/2}
    )^\mathsf{T},
  \end{aligned}
\end{align}
whereas for $V=\varepsilon V_-$ they become
\begin{align}
  \begin{aligned}
    \relax[\bra{\tilde a_{2,3}(\varepsilon)}]_L
                            & \sim
    \mqty(
    -\frac{1}{3}\varepsilon &
    1                       &
    \pm \frac{i}{\sqrt{3}}\varepsilon^{1/2}
    ),
    \\
    [\ket{a_{2,3}(\varepsilon)}]_L
                            & \sim
    \mqty(
    -\frac{1}{3}\varepsilon                 &
    \pm \frac{i}{\sqrt{3}}\varepsilon^{1/2} &
    1
    )^\mathsf{T}.
  \end{aligned}
\end{align}
Thus, in the limit $\varepsilon\to0$, the right eigenvectors $\ket{a_{2}(\varepsilon)}$ and $\ket{a_{3}(\varepsilon)}$ satisfy
% \begin{align}
%   \lim_{\varepsilon\to0}\ket{a_{2,3}(\varepsilon)} & =\ket{L_+}
%   \quad
%   (V=\varepsilon V_+),
%   \\
%   \lim_{\varepsilon\to0}\ket{a_{2,3}(\varepsilon)} & =\ket{L_-}
%   \quad
%   (V=\varepsilon V_-),
% \end{align}
\begin{align}
  \lim_{\varepsilon\to0}\ket{a_{2,3}(\varepsilon)} & =\ket{L_\pm}
  \quad
  (V=\varepsilon V_\pm),
\end{align}
demonstrating explicit coalescence onto one of the two chiral modes.

At exactly $\varepsilon=0$, however, the characteristic equation reduces to $(\lambda+1)^2(\lambda-2)=0$, and the eigenspace at $\lambda=-1$ is spanned by $\ket{L_+}$ and $\ket{L_-}$. The unperturbed Hamiltonian $H_0$ is therefore diagonalizable, because one may freely choose orthogonal linear combinations
% \[
%   \ket{a_{2,3}}=c_1^{(2,3)}\ket{L_+}+c_2^{(2,3)}\ket{L_-}
% \]
within the degenerate subspace so as to satisfy $\braket{a_2}{a_3}=0$. By contrast, for {{infinitesimally small but nonzero}} $\varepsilon\neq0$, the eigenstates approach a maximally nonorthogonal configuration through coalescence onto a single chiral mode. This is the essential EP-like feature of the mechanism. Indeed, the phase rigidity obeys the leading-order scaling $r_j\propto \varepsilon^{1/2}$, identical to the standard exponent at an ordinary EP, and vanishes in the limit $\varepsilon\to0$, indicating self-orthogonality.

The origin of this unusual behavior becomes transparent upon constructing an effective $2\times2$ Hamiltonian in the degenerate subspace
\begin{align}
  % \[
  P=\ket{L_+}\bra{L_+}+\ket{L_-}\bra{L_-}.
  % \]
\end{align}
Letting $Q=I-P$ denote the complementary subspace, the Feshbach projection formalism gives \cite{feshbach-1958-AnnualReviewofNuclearandParticleScience-OpticalModelIts}
\begin{align}
  H_{\mathrm{eff}}(\lambda)
  =
  PHP + PVQ\frac{1}{\lambda-QHQ}QVP.
  \label{eq:rigorous_Feshbach-Heff}
\end{align}

{{For $V=\varepsilon V_+$, expanding $H_{\mathrm{eff}}$ in powers of $\varepsilon$
and retaining the lowest nontrivial orders, we obtain
\begin{align}
  [H_{\mathrm{eff}}^{(+)}]_L
  =
  \mqty(
  -1             & \varepsilon \\
  -\varepsilon^2/3 & -1
  )
  + O(\varepsilon^3),
  \label{eq:Heff_plus}
\end{align}
where the derivation is given in Appendix~\ref{app:A}. Likewise, for $V=\varepsilon V_-$ we similarly find
\begin{align}
  [H_{\mathrm{eff}}^{(-)}]_L
  =
  \mqty(
  -1               & -\varepsilon^2/3 \\
  \varepsilon      & -1
  )
  + O(\varepsilon^3).
  \label{eq:Heff_minus}
\end{align}
Equations~\eqref{eq:Heff_plus} and \eqref{eq:Heff_minus} show that one coupling appears already at
$O(\varepsilon)$, while the reverse coupling is suppressed to $O(\varepsilon^2)$.
Thus, for $V=\varepsilon V_\pm$ the effective coupling from the $\ket{L_\mp}$
component to the $\ket{L_\pm}$ component appears at $O(\varepsilon)$, whereas the
reverse process from $\ket{L_\pm}$ to $\ket{L_\mp}$ arises only at
$O(\varepsilon^2)$ via the intermediate state $\ket{L_0}$.}}

Accordingly, in the limit $\varepsilon\to0$, the leading part of the effective
Hamiltonian approaches a Jordan form only asymptotically:
\begin{align}
  \begin{aligned}
      [H_{\mathrm{eff}}^{(+)}]_L
  &\sim
  \mqty(
  -1 & \varepsilon \\
  0 & -1
  ), &&
  [H_{\mathrm{eff}}^{(-)}]_L
  &\sim
  \mqty(
  -1 & 0 \\
  \varepsilon & -1
  ).
  \end{aligned}
\end{align}
This explains why both eigenvectors asymptotically coalesce onto $\ket{L_+}$ for
$V=\varepsilon V_+$ and onto $\ket{L_-}$ for $V=\varepsilon V_-$. At the same time, because the two off-diagonal terms scale as $O(\varepsilon)$ and $O(\varepsilon^2)$, the eigenvalue splitting is $\Delta\lambda \sim \sqrt{\varepsilon\cdot\varepsilon^2} = \varepsilon^{3/2}$, reproducing Eq.~\eqref{eq:3mode_eigenvalue_perturbation}. The exponent $3/2$ therefore originates from the coexistence of a first-order direct coupling and a second-order indirect coupling mediated by the nondegenerate mode $\ket{L_0}$.

For comparison, we consider the standard $2\times2$ representation of an ordinary EP \cite{kato-1995--PerturbationTheoryLinear, moro-1997-SIAMJournalonMatrixAnalysisandApplications-LidskiiVishikLyusternikPerturbationTheory} :
\begin{align}
  H_{\mathrm{EP}}
  = \mqty(
  \lambda_{\mathrm{EP}} & \varepsilon_A \\
  \varepsilon_B & \lambda_{\mathrm{EP}}
  ).
\end{align}
In that case, the eigenvalue splitting scales as $\Delta\lambda \sim \sqrt{\varepsilon_A\varepsilon_B}$, and at the EP itself one of the two effective couplings vanishes at a finite operating point, so that the Hamiltonian becomes exactly Jordan-block defective. The usual square-root response $\Delta\lambda\propto\varepsilon^{1/2}$ is a direct consequence of this Jordan-block structure.

{ This observation motivates a more general characterization of the { AEP}. We may summarize the AEP response by the effective two-mode form
\begin{align}
  \begin{aligned}
      H_{\mathrm{AEP}}
  \sim
  \mqty(
    \lambda_{\mathrm{DP}} & c_1 \varepsilon^p \\
    c_2 \varepsilon^q & \lambda_{\mathrm{DP}}
  ),
  \end{aligned}\label{eq:AEP_general}
\end{align}
where $c_1, c_2 \neq 0$ and $p, q$ are positive integers. The essential feature of the AEP is the perturbative asymmetry $p \neq q$. Under this condition, the eigenvalue splitting follows the power law $\Delta\lambda \sim \varepsilon^{(p+q)/2}$, while the eigenvectors approach a Jordan-type defective structure only asymptotically in the limit $\varepsilon\to0$. {  
This effective-Hamiltonian form makes explicit the asymmetric perturbative coupling structure responsible for the AEP response.} The same asymmetric coupling mechanism is not restricted to the minimal three-resonator model. When the phase-circulating perturbation $V_\pm$ introduced in Eq.~\eqref{eq:3mode_perturbation_H_pm} is extended to a cyclic $N$-site system, the same perturbative asymmetry generally persists in the effective Hamiltonian for each degenerate OAM pair and can therefore induce chiral-mode selection also for higher-order OAM doublets. This generalization is discussed in Appendix~\ref{APP:generalization_high_dimensional}.}

{ We have thus established { the first switching principle organized by
the AEP:} switching from an achiral state to a chiral state through chiral-mode selection induced by an infinitesimal complex perturbation. }

{
\section{CHIRALITY REVERSAL VIA AN EP PAIR IN THE VICINITY OF THE AEP}\label{sec3:chiral_switching}}

{
In Sec.~\ref{sec2:chiral_mode_emergence}, we established { the first switching principle: switching from an achiral state to a chiral state via the direct AEP response.} We now turn to the second principle and ask whether the chirality generated { by this AEP response} can be reversed. To address this question, we introduce a two-parameter non-Hermitian perturbation. { As shown below, { this enlarged parameter space contains exceptional-line branches of ordinary EPs that meet at the AEP.} Chirality reversal is realized by a one-dimensional control sweep performed at a finite bias away from the AEP, which crosses the two branches associated with opposite chiralities.

This geometric relation also has an operational consequence. { If one intends to reduce the perturbation strength required for EP-based chirality reversal, the two EPs should be placed as close as possible to the AEP.} Therefore, the small-{ perturbation} limit of EP-pair switching is asymptotically connected to the direct AEP response.}}

\subsection{Two-parameter non-Hermitian perturbation}

\begin{figure*}[tbph]
  \centering
  \includegraphics[width=\linewidth]{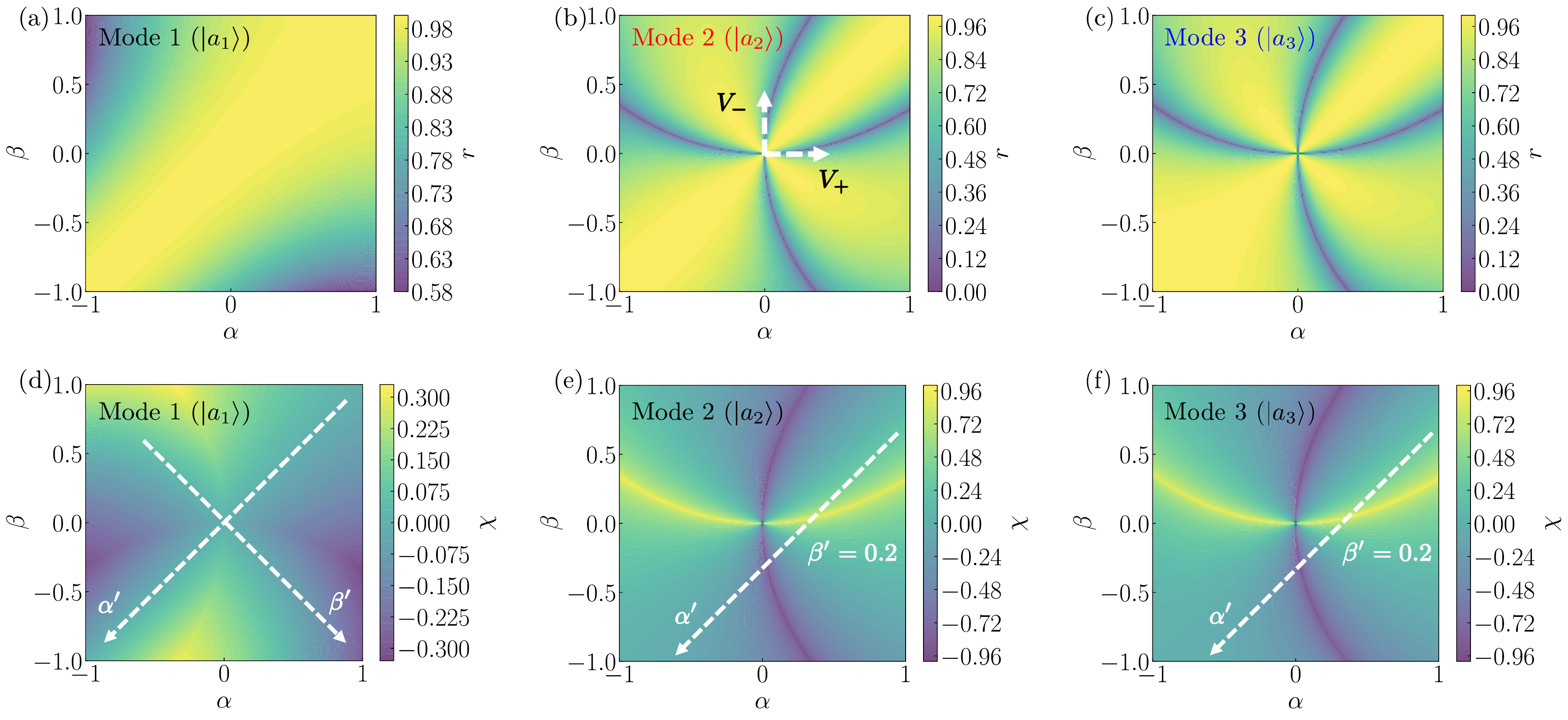}
\caption{Results for $\theta=0$. (a)--(c) Phase rigidity $r_j$ and (d)--(f) chirality $\chi_j$ as functions of $\alpha$ and $\beta$ for modes $j=1, 2, 3$. { White dashed arrows in (b) on the $\beta=0$ and $\alpha=0$ axes indicate the $V_+$- and $V_-$-induced AEP perturbation directions identified in Sec.~II.} For $j=2,3$ in (b, c) and (e, f), EPs appear along specific parameter curves (bright and dark lines) where eigenvectors coalesce and chirality reaches extremal values $\chi_j=\pm1$. The origin $(\alpha,\beta)=(0,0)$ corresponds { to the AEP point}, where these curves intersect. Similar arrows in (d) indicate the $\alpha'$ and $\beta'$ axes for the perturbation in Eq.~\eqref{eq:3mode_hamiltonian_general_rotation} at $\theta=3\pi/4$, while those in (e) and (f) mark the $\alpha'$ axes at $\beta' = 0.2$.}
  \label{fig:3mode_chiral}
\end{figure*}

\begin{figure*}[htbp]
  \centering
  \includegraphics[width=\linewidth]{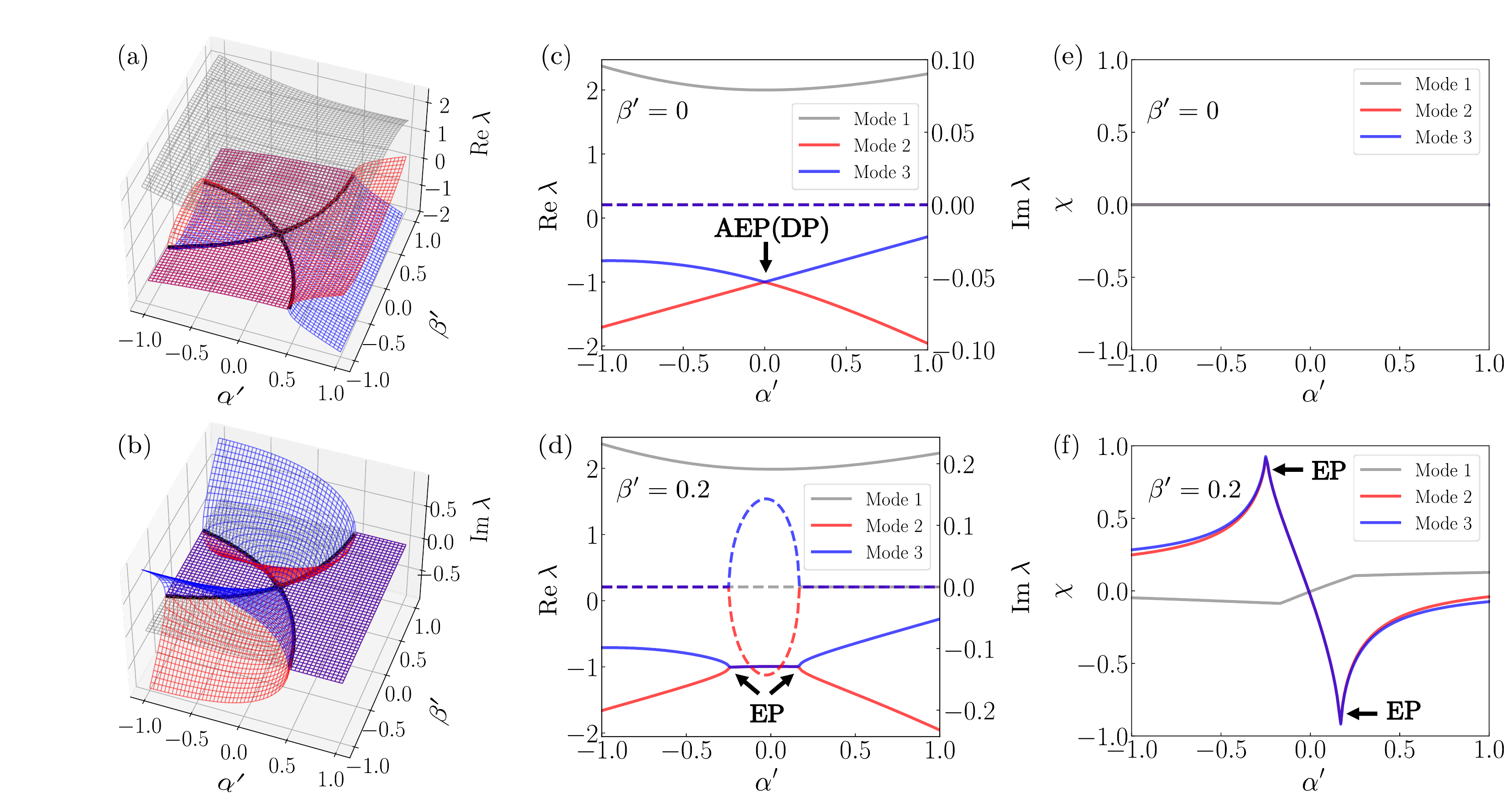}
  \caption{Eigenvalue structure at $\theta=3\pi/4$. (a), (b) Eigenvalues $\lambda_j$ as functions of $\alpha'$ and $\beta'$ (gold: $j=1$, red: $j=2$, blue: $j=3$). {{The red curves indicate the {{exceptional lines}}, along which the eigenvalue surfaces of $\lambda_2$ and $\lambda_3$ coalesce. 
  }} 
  (c), (d) Eigenvalue variation as a function of $\alpha'$ for fixed $\beta'=0$ and $\beta'=0.2$, respectively. For $\beta'\neq0$, { the $\alpha'$ scan crosses the two exceptional-line branches.} (e), (f) Chirality as a function of $\alpha'$ for fixed $\beta'=0$ and $\beta'=0.2$, respectively. For $\beta'=0$, no chirality appears, whereas a nonzero imaginary bias $\beta'$ enables switching between the $\chi=\pm1$ states associated with the EPs by tuning the real perturbation $\alpha'$ alone.}
  \label{fig:3mode_chiral_135_combined}
\end{figure*}

We introduce a model Hamiltonian based on two non-Hermitian perturbations, $V^{(1)}(\theta)$ and $V^{(2)}(\theta)$. These perturbations extend the single-channel perturbations $V_+$ and $V_-$ introduced in Sec.~\ref{sec2:chiral_mode_emergence}. They are defined as 
\begin{gather}
  H(\theta) = H_0 + \alpha(\theta) V^{(1)}(\theta) + \beta(\theta) V^{(2)}(\theta),
  \label{eq:3mode_hamiltonian_general_rotation}
  \\
  V^{(1)}(\theta)= \cos\theta \, V_+ - \sin\theta \, V_-,
  \label{eq:3mode_perturbation_H1}
  \\
  V^{(2)}(\theta)= \sin\theta \, V_+ + \cos\theta \, V_-.
  \label{eq:3mode_perturbation_H2}
\end{gather}
Here, $\alpha(\theta)$ and $\beta(\theta)$ denote the perturbation amplitudes of $V^{(1)}$ and $V^{(2)}$, normalized by the coupling {{rate}}. For brevity, we denote the perturbation amplitudes at $\theta=0$ by $(\alpha,\beta):=(\alpha(0),\beta(0))$. The chirality of the eigenstates $\ket{a_j}$ ($j=1,2,3$) of this Hamiltonian is evaluated using Eq.~\eqref{eq:chirality_definition}. At $\theta=0$, the two coordinate axes in the $(\alpha,\beta)$ plane coincide with the single-channel perturbation directions studied in Sec.~II: the $\beta=0$ axis corresponds to the $V_+$ direction and the $\alpha=0$ axis to the $V_-$ direction. { These two axes therefore represent the AEP perturbation directions in the present two-parameter space.}

{{To visualize the dependence of the eigenstate properties on the two perturbation parameters, we map the phase rigidity and chirality in the $(\alpha, \beta)$ plane. As seen in Figs.~\ref{fig:3mode_chiral}(a) and \ref{fig:3mode_chiral}(d), the nondegenerate mode $\ket{a_1}$ does not participate in the coalescence of the chiral pair. Accordingly, its phase rigidity does not vanish, and its chirality remains close to zero throughout the parameter space. By contrast, Figs.~\ref{fig:3mode_chiral}(b) and \ref{fig:3mode_chiral}(c) reveal that {two one-dimensional loci in the $(\alpha,\beta)$ plane along which the phase rigidity vanishes and the two chiral modes coalesce. These loci are therefore ordinary EPs, which we refer to as exceptional lines.} As shown in Figs.~\ref{fig:3mode_chiral}(e) and \ref{fig:3mode_chiral}(f), these {{exceptional lines}} coincide with extrema of the chirality, $\chi=\pm1$. 
}}

Importantly,  the strongest chiral response associated with the AEP is obtained along the coordinate axes, specifically the $V_+$ and $V_-$ directions. As seen in Fig.~\ref{fig:3mode_chiral}(b), { ordinary EPs lie on exceptional-line branches that meet at the AEP. The $V_+$ and $V_-$ AEP perturbation directions coincide with the tangent directions of these branches at the AEP. Thus, within the same parameter space, the two switching principles are geometrically connected.} 

This geometric relation clarifies that the chirality switching discussed here has two distinct but related operating principles. One is the AEP-based principle established in Sec.~II, where the system is switched from an achiral state to a chiral state { along the $V_\pm$ AEP perturbation directions.} The other is the EP-based principle, in which the system is switched between opposite chiral states by crossing the exceptional lines.

\subsection{Separated real and imaginary control axes}

{ We next identify a practically useful set of control axes { for implementing the second switching principle revealed above, namely, chirality reversal by crossing the exceptional lines in parameter space. The key point is that this operation can be realized by moving the system between two ordinary EPs associated with opposite chiralities.} We therefore seek a parameterization in which this motion can be implemented by independently controlled real and imaginary perturbation components.}

Changing $\theta$ rotates the parameter-space structure shown in Fig.~\ref{fig:3mode_chiral}, thereby allowing different choices of control axes associated with $V^{(1)}$ and $V^{(2)}$. In particular, when $\theta=(2n-1)\pi/4$ $(n=1,2,3,4)$, $V^{(1)}$ and $V^{(2)}$ can be chosen so that one corresponds to a purely real perturbation and the other to a purely imaginary perturbation. This separation is advantageous in realistic implementations, because independently tuning real or imaginary perturbations is generally much easier than controlling both components simultaneously. In particular, for $\theta=3\pi/4$, the perturbation takes the form
\begin{align}
  \begin{aligned}
       & \qty[V\qty(\frac{3\pi}{4})]_{\mathrm{site}}   = \sqrt{2}
    \alpha'
    \mqty(
    -1 & 0                                                        & 0           \\
    0  & \frac{1}{2}                                              & 0           \\
    0  & 0                                                        & \frac{1}{2}
    )
    +
    \sqrt{\frac{3}{2}} \beta'
    \mqty(
    0  & 0                                                        & 0           \\
    0  & -i                                                       & 0           \\
    0  & 0                                                        & i
    )
  \end{aligned}
  \label{eq:perturbation_135_local}
\end{align}
where $(\alpha',\beta')=(\alpha(3\pi/4),\beta(3\pi/4))$. The corresponding $\alpha'$ and $\beta'$ axes are shown by the white dashed lines in Figs.~\ref{fig:3mode_chiral}(d).

{{
  This perturbation matrix provides a decomposition of the perturbation into real and imaginary onsite components. For the discussion below, we adopt a representative control scenario in which the real perturbation $\alpha'$ is used as the dynamically tuned parameter, while the imaginary perturbation $\beta'$ provides a fixed loss bias. While this assignment is chosen for concreteness, we note that both components may be tunable depending on the platform.
  }} 
In this way, the system can be moved between the bright line ($\chi=1$) and the dark line ($\chi=-1$) in Figs.~\ref{fig:3mode_chiral}(e) and \ref{fig:3mode_chiral}(f). {
  This separation of the control axes is also advantageous from the viewpoint of implementation. In practice, the imaginary component may be provided by spatially selective gain or loss engineering \cite{nomura-2009-OpticsExpress-PhotonicCrystalNanocavity,ellis-2011-NaturePhotonics-UltralowthresholdElectricallyPumped,takata-2017-PhysicalReviewApplied-$mathcalPmathcalT$SymmetricCoupledResonatorWaveguide}, while the real component may be controlled through resonance-frequency tuning mechanisms such as thermo-optic \cite{brunstein-2009-OpticsExpress-ThermoopticalDynamicsOptically,notomi-2008-AdvancesinOpticalTechnologies-OnChipAllOpticalSwitching}, carrier-injection \cite{tanabe-2009-OpticsExpress-LowPowerFast,iadanza-2020-PhysicalReviewB-ModelThermoopticNonlinear}, or post-fabrication local tuning and non-volatile phase-change loading of photonic-crystal nanocavities \cite{chen-2011-OpticsExpress-SelectiveTuningHighQ,intonti-2012-AppliedPhysicsLetters-ModeTuningPhotonic,uemura-2024-ScienceAdvances-PhotonicTopologicalPhase}. In this sense, the separated-control formulation provides a realistic route to EP-based chirality switching without requiring direct tuning of a fully complex perturbation.} 

Figure~\ref{fig:3mode_chiral_135_combined} summarizes how this switching appears in the eigenvalue structure and chirality response. Figs.~\ref{fig:3mode_chiral_135_combined}(a) and \ref{fig:3mode_chiral_135_combined}(b) show the two-parameter eigenvalue structure, where the red curves indicate the {{exceptional lines}} along which the eigenvalue surfaces of $\lambda_2$ and $\lambda_3$ coalesce. The one-dimensional cuts in Figs.~\ref{fig:3mode_chiral_135_combined}(c) and \ref{fig:3mode_chiral_135_combined}(d) show that, for $\beta'= 0$, the degeneracy at the DP is lifted continuously as $\alpha'$ is varied. In this case, the system does not cross any exceptional line, and no chirality emerges. By contrast, for $\beta'\neq 0$, the $\alpha'$ scan crosses the {{exceptional lines}}. The corresponding chirality response is shown in Figs.~\ref{fig:3mode_chiral_135_combined}(e) and \ref{fig:3mode_chiral_135_combined}(f): for $\beta'=0$, the chirality remains zero, while for $\beta'\neq0$ it switches between $\chi=\pm1$ as $\alpha'$ is swept across the two EPs. 
  
{ In this representation, { the second switching principle, namely chirality reversal between opposite chiral states,} is realized not by following the { $V_\pm$ AEP perturbation directions} themselves, but by crossing the exceptional lines between opposite-chirality regions through separated real and imaginary controls.}

\subsection{Effective Hamiltonian and switching condition in the degenerate chiral subspace}

{ To quantify {the EP-based chirality-reversal condition near the AEP}, we now derive the effective Hamiltonian in the degenerate chiral subspace for $H(3\pi/4)$. This reduced description makes it possible to extract a simple relation between the imaginary bias $\beta'$ and the real control amplitude $\Delta\alpha'$ required for chirality switching.} {It also clarifies how the EP-based switching condition {is connected to the local geometry of the exceptional-line branches near the AEP,} even though the switching points themselves are ordinary EPs { located at nonzero parameter values on these branches.}} {{The effective Hamiltonian in {{the degenerate chiral-mode subspace}} for $H(3\pi/4)$ is given by}}
\begin{align}
  [H_{\mathrm{eff}}]_L \simeq
  \mqty(
  -1 + \frac{\beta'^2 - \alpha'^2}{6}
   & \frac{\beta'-\alpha'}{\sqrt{2}} - \frac{(\beta' + \alpha')^2}{6} \\
  -\frac{\beta' + \alpha'}{\sqrt{2}} - \frac{(\beta' - \alpha')^2}{6}
   & -1 + \frac{\beta'^2 - \alpha'^2}{6}
  ).
  \label{eq:3mode_perturbation:Heff135}
\end{align}
Defining
\begin{align}
  A & = \frac{\beta'-\alpha'}{\sqrt{2}} - \frac{(\beta' + \alpha')^2}{6},    \\
  B & = -\frac{\beta' + \alpha'}{\sqrt{2}} - \frac{(\beta' - \alpha')^2}{6}.
\end{align}
% \begin{align}
%   \begin{aligned}
%       A = \frac{\beta'-\alpha'}{\sqrt{2}} - \frac{(\beta' + \alpha')^2}{6},    &&
%   B = -\frac{\beta' + \alpha'}{\sqrt{2}} - \frac{(\beta' - \alpha')^2}{6},
%   \end{aligned}
% \end{align}
{ Unlike the AEP effective Hamiltonian in Eq.~\eqref{eq:AEP_general}, whose essential feature is a perturbative asymmetry between the two directional couplings, the present effective Hamiltonian describes { ordinary EPs located at nonzero parameter values on the exceptional-line branches. These EP conditions are given by } the vanishing of one of the two effective couplings, $A=0$ or $B=0$. The switching points found here are therefore ordinary EPs rather than asymptotic ones, and the conditions $A=0$ and $B=0$ trace the exceptional lines.} In the weak-perturbation regime ($|\alpha'|\ll 1, |\beta'| \ll 1$), the real perturbation required to reverse the chirality from $\chi=\pm1$ to $\chi=\mp1$, i.e., to move between the two EPs, is therefore
\begin{align}
  \Delta \alpha' = 2\beta' + o(\beta').
\end{align}
This relation shows that, { if one intends to reduce the perturbation strength required for EP-based chirality reversal, the two EPs should be placed as close as possible to the AEP}. In the ideal perturbative limit, this corresponds to taking $|\beta'|$ as small as possible, so that the two EPs approach the { AEP} while the required real control amplitude is simultaneously reduced. { This limit is precisely the regime in which the exceptional lines are locally described by their tangent directions at the AEP, namely the $V_+$ and $V_-$ AEP perturbation directions identified in Fig.~\ref{fig:3mode_chiral}. Thus, in the low-perturbation limit, the EP-pair route becomes asymptotically connected to the direct AEP route through the same local exceptional-line geometry.}
{
In practice, this limiting procedure is constrained by the finite resolution of parameter control. We return to this point in the next section, where the direct AEP and EP-pair operating points are compared under a finite-resolution model. As discussed in Appendix~\ref{APP:generalization_high_dimensional}, this EP-based structure also persists in cyclic $N$-site systems.}

\section{Discussion}
\label{sec:discussion}

{
  \subsection{Two distinct chirality-switching principles organized by the AEP}}

{ The central conclusion of the preceding sections is that { the AEP organizes two related operating principles of chirality switching. One is the direct AEP response}, in which an infinitesimal complex perturbation induces chiral-mode selection and thereby switches the system from an achiral state to a chiral state. The other is the EP-based principle {in the vicinity of the AEP}, in which ordinary EPs { located at nonzero parameter values on exceptional-line branches} enable chirality reversal between opposite chiral states. 

{Importantly, this EP pair is not an independent singular structure separated from the AEP. { As shown in Fig.~\ref{fig:3mode_chiral}, ordinary EPs lie on exceptional-line branches that meet at the AEP, and the tangent directions of these branches at that point coincide with the AEP perturbation directions. Therefore, to achieve EP-based chirality reversal with smaller perturbation amplitudes, the relevant EPs move closer to the AEP along these exceptional-line branches. In this sense, the EP-pair route is asymptotically connected to the direct AEP-based switching route in the low-perturbation limit.}} The purpose of the following discussion is therefore to compare these two { AEP-organized operating routes} under finite-resolution control and to clarify their different { practical implications for chiral-mode control.}}

\begin{figure*}[tbph]
  \centering
  \includegraphics[width=\linewidth]{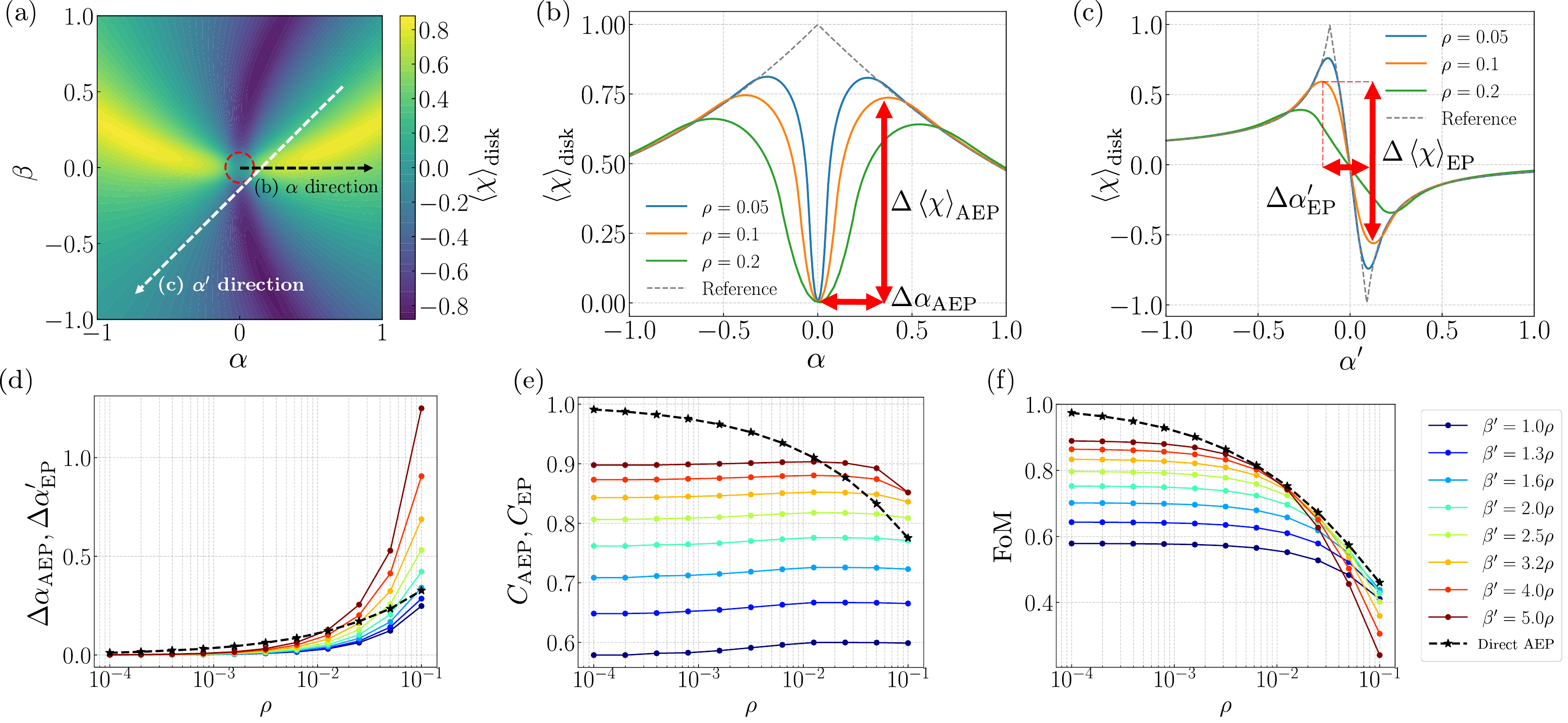}
  \caption{Finite-resolution comparison { of the direct AEP operating point and
the EP-pair operating point.} (a) Chirality map in the control-parameter plane for $\rho=0.1$. { (b) Disk-averaged chirality $\ev{\chi}_{\mathrm{disk}}$ as a function of
$\alpha$ for the AEP-based switching scheme, shown for $\rho=0.05$, $0.1$,
and $0.2$. The gray dashed curve shows the reference result obtained
without resolution averaging. The red arrows indicate the chirality
change $\Delta\ev{\chi}_{\mathrm{AEP}}$ and the corresponding real control
excursion $\Delta\alpha_{\mathrm{AEP}}$.
(c) Disk-averaged chirality $\ev{\chi}_{\mathrm{disk}}$ as a function of
$\alpha'$ for the EP-based switching scheme at $\beta'=0.1$, shown for
$\rho=0.05$, $0.1$, and $0.2$. The gray dashed curve shows the reference
result obtained without resolution averaging. The red arrows indicate the
total chirality change $\Delta\ev{\chi}_{\mathrm{EP}}$ and the
corresponding real control excursion $\Delta\alpha'_{\mathrm{EP}}$.} (d) Perturbation amplitude { required to realize chirality switching.} (e) { Normalized chirality-change ratio} as a function of $\rho$, with
$C_{\mathrm{AEP}}=\Delta\ev{\chi}_{\mathrm{AEP}}$ and
$C_{\mathrm{EP}}=\Delta\ev{\chi}_{\mathrm{EP}}/2$. (f) Engineering figure of merit defined in Eq.~\eqref{eq:fom} { for operating from a chirality-neutral state.} In the sufficiently high-resolution regime for $\rho \lesssim 10^{-3}$, the { direct AEP} operating point can outperform the { EP-pair operating point} within the present averaging model. { A complementary single-branch achiral-to-chiral FoM for the EP branch is presented in Appendix~\ref{app:alternatives_gene}.}}
  \label{fig:rhoboke}
\end{figure*}

{
\subsection{Resolution-averaged chirality near the AEP}}

{
The comparison above suggests that the key practical issue is not merely the existence of {  two AEP-related singular responses}, but how they should be interpreted as experimentally accessible operating points. 
}
The chiral response associated with the {{AEP}}, discussed in Sec.~\ref{sec2:chiral_mode_emergence}, is defined theoretically in the infinitesimal-perturbation limit $\varepsilon \to 0$. To interpret this response as a realistic device operating point, one must therefore reconsider { how the AEP can be accessed in an experimentally controlled parameter space.} In the present $(\alpha(\theta),\beta(\theta))$ parameterization, the { AEP} is an isolated point in a two-dimensional control space. Although it is well defined within the idealized theoretical model, reaching it exactly cannot be guaranteed in practice because of finite control resolution, drift, noise, and calibration errors.

From this perspective, experimentally observed quantities should be interpreted not as values at a single ideal point, but as responses averaged over a finite region in control-parameter space. We therefore model the finite control resolution in the perturbation parameters $(\alpha,\beta)$ at $\theta = 0$, by a uniform distribution over a disk of radius $\rho$, and define the observable chirality as
\begin{gather}
  \begin{aligned}
     & \expval{\chi}_{\mathrm{disk}}(\alpha,\beta) \\
     & \quad =
    \iint
    \chi(\alpha+\Delta\alpha,\beta+\Delta\beta)\,
    p_{\mathrm{disk}}(\Delta\alpha,\Delta\beta)\,
    d\Delta\alpha\, d\Delta\beta, \notag
  \end{aligned}
  \\
  p_{\mathrm{disk}}(\Delta\alpha,\Delta\beta)
  =
  \frac{1}{\pi \rho^2}
  \mathbf{1}\!\left(\Delta\alpha^2+\Delta\beta^2\le \rho^2\right).
\end{gather}
Such averaging provides a natural way to reinterpret the response near the { AEP} in terms of experimentally accessible observables. Here we adopt the disk average as the minimal isotropic model of finite control resolution in the $(\alpha,\beta)$ plane. As shown in Appendix~\ref{app:avg_fom}, however, the qualitative comparison between the two operating points remains unchanged if one instead assumes one-dimensional averaging only along the imaginary (loss) direction.

Under this averaging, the { AEP-based operating point} appears fundamentally differently from its idealized limit. In principle, an infinitesimal perturbation along the $V_\pm$ direction yields a maximally chiral response. 
Once a finite resolution $\rho$ is introduced, however, { the exact AEP at $\varepsilon=0$ is averaged over the crossing of positive- and negative-chirality regions and therefore appears effectively as a neutral point with $\chi=0$.} As a consequence, the { idealized achiral-to-chiral switching response}, in which $\chi$ approaches $\chi=-1$ or $1$ for arbitrarily small perturbations, is smoothed into a finite-width operating region. { To quantify this averaged AEP-based response, we use Fig.~\ref{fig:rhoboke}(b). There, we define $\Delta\ev{\chi}_{\mathrm{AEP}}$ as the change in the resolution-averaged chirality from the chirality-neutral point to the maximum value reached along the { $V_+$ AEP direction.} We also define $\Delta\alpha_{\mathrm{AEP}}$ as the corresponding control excursion required to reach that maximum averaged chirality, as indicated by the red arrows in Fig.~\ref{fig:rhoboke}(b).}

A similar limitation arises for EP-based chirality switching. In Sec.~\ref{sec3:chiral_switching}, we showed that sweeping $\alpha'$ at $\beta'\neq0$ allows chirality switching via EPs. When $\beta' \lesssim \rho$, however, the averaging strongly mixes the neighboring positive- and negative-chirality regions, making it difficult to retain a large chirality contrast. Thus, although EP switching remains operable under finite resolution, operation with an arbitrarily small loss bias $\beta' \to 0$ is not realistic in the presence of uncertainty; instead, an effective lower bound of order $\beta' \gtrsim \rho$ emerges. { To quantify the averaged EP-based response, we use Fig.~\ref{fig:rhoboke}(c). For a fixed loss bias $\beta'$, we define $\Delta\ev{\chi}_{\mathrm{EP}}$ as the total change in the
resolution-averaged chirality between the two opposite-chirality extrema along the $\alpha'$ sweep. We also define $\Delta\alpha'_{\mathrm{EP}}$ as the corresponding control excursion
required to connect these extrema, again as indicated by the red arrows in Fig.~\ref{fig:rhoboke}(c).} {{These qualitative differences are illustrated in Figs.~\ref{fig:rhoboke}(a)--(c). Fig.~\ref{fig:rhoboke}(a) shows the chirality map in the $(\alpha,\beta)$ plane together with the representative parameter-sweep directions, while Figs.~\ref{fig:rhoboke}(b) and (c) compare the { direct AEP and EP-pair operating points} under finite resolution. }} 

In this way, the two { AEP-organized switching} principles are both degraded by finite control resolution, but in different ways. To quantify these differences, we now introduce a figure of merit.

\subsection{Comparison via a finite-resolution figure of merit}
{
As the above discussion makes clear, the comparison between AEP- and
EP-based operating points cannot be assessed solely from the raw
resolution-averaged chirality change, because the two switching processes
have different ideal chirality changes. The AEP-based process switches
from a chirality-neutral state to a chiral state, whose ideal chirality
change is unity, whereas the EP-based process switches between two
opposite chiral states, whose ideal chirality change is two. We therefore
introduce the normalized chirality-change ratio
\begin{align}
  \begin{aligned}
      C_{\mathrm{AEP}} (\rho)= \Delta \ev{\chi}_{\mathrm{AEP}}, &&
  C_{\mathrm{EP}}(\rho) = {\Delta \ev{\chi}_{\mathrm{EP}}}/{2},
  \end{aligned} \label{eq:chirality_change_ratio}
\end{align}
 With this normalization, $C=1$ represents the ideal chirality change for each switching process, and the reduction of $C$ from unity directly quantifies the degradation caused by finite control resolution.}

Beyond the difference in the ideal chirality changes discussed above, the
two switching processes also differ in the structure of the perturbations
required for their implementation. In the AEP-based case, the control
excursion $\Delta\alpha_{\mathrm{AEP}}$ is associated with a complex
phase-circulating perturbation along the { $V_+$ AEP direction}. Thus, the
chirality change is produced by controlling a complex onsite modulation
itself. In contrast, in the EP-based case, the switching excursion
$\Delta\alpha'_{\mathrm{EP}}$ can be implemented by a real onsite
perturbation, but only in the presence of a nonzero imaginary bias
$\beta'$ that is applied in advance.
Therefore, a fair device-oriented comparison should evaluate not only the
normalized chirality-change ratio $C$, but also the real-control excursion
required for switching and the gain/loss component required to set the
operating condition. We therefore introduce the performance metric
\begin{align}
  \mathrm{FoM}(\rho)
  =
  \frac{C(\rho)}
  {(1+\Delta(\rho))(1+\Gamma(\rho))}{{.}} \label{eq:fom}
\end{align}
{{Here, $\Delta(\rho)$ and $\Gamma(\rho)$ should be understood as the real-control cost and the gain/loss cost, respectively, evaluated at the operating point that { realizes chirality switching under resolution-averaged conditions}; the explicit operational definitions used for the AEP- and EP-based cases are given in Appendix~\ref{app:fom_explicit}.}}
This quantity is an engineering metric designed to compare response amplitude, control cost, and loss cost on equal footing.

Within this metric, { the direct AEP operating point and the EP-pair operating point} {  exhibit different finite-resolution performance.} 
{{The quantitative comparison is summarized in Figs.~\ref{fig:rhoboke}(d)--(f). Fig.~\ref{fig:rhoboke}(d) shows the perturbation amplitude required to realize the corresponding chirality switching $\Delta \alpha_{\mathrm{AEP}}, \Delta \alpha'_{\mathrm{EP}}$, while Fig.~\ref{fig:rhoboke}(e) shows the normalized chirality-change ratio $C$ as a function of the control resolution $\rho$. Fig.~\ref{fig:rhoboke}(f) combines these factors through the figure of merit.}}

{ An important point is that the high-resolution regime is precisely the regime in which the EP-pair operating point { lies} close to the AEP. { In this limit, the EP-based operation uses ordinary EPs on exceptional lines whose local tangent directions at the AEP coincide with the $V_\pm$ AEP perturbation directions.} Nevertheless{ , even in this limit where the operating EPs lie close to the AEP}, the direct AEP operating point can exhibit a higher figure of merit within the present averaging model. }

Because this comparison depends on the assumed averaging model and cost function, the present FoM analysis should be understood as a design guideline rather than a universal claim.

{ In the main text, we evaluate the EP-based operating point in its natural role as a full chirality-reversal process between opposite chiral states. For completeness, however, Appendix~\ref{app:alternatives_gene} also considers a complementary single-branch perturbation-sweep interpretation of the same EP branch, in which the response is evaluated from a chirality-neutral reference point to one EP-associated chiral extremum. This alternative comparison is not used for the main-text FoM, but it provides a task-matched reference to the AEP-based achiral-to-chiral switching and leads to the same qualitative conclusion in the sufficiently high-resolution regime.}

\subsection{Scope and limitations}

Our treatment is based on { temporal coupled-mode theory} and effective Hamiltonians, and is intended primarily to clarify the conditions under which the two { chirality} switching principles arise, together with the basic response laws that characterize them. Accordingly, several important effects lie beyond the scope of the present study, including validation by full-wave electromagnetic simulations, statistical effects of fabrication disorder, gain saturation, nonlinear response, and time-dependent noise. In addition, the chirality employed here is defined through projection onto an OAM basis, and its direct one-to-one correspondence with experimentally measurable observables remains to be established for specific physical implementations. {
These limitations do not alter the main conceptual result: {AEP-based achiral-to-chiral switching and EP-pair-based chiral-to-chiral switching define two distinct operating principles organized by the same AEP.}}

\section{Conclusion}

In this work, we developed a theoretical framework for chiral-mode control in discrete coupled single-mode resonators. Using the minimal three-resonator model, we showed that a phase-circulating complex onsite perturbation { defines an asymptotic exceptional point (AEP): { a Hermitian DP whose effective two-mode Hamiltonian acquires asymmetric perturbative coupling orders and whose eigenvectors coalesce asymptotically as the perturbation strength tends to zero.}} We then showed that, when the perturbation space is enlarged, { ordinary EPs lie on exceptional-line branches that meet at the AEP. A finite-bias control sweep crosses these branches at an EP pair, enabling chirality reversal between opposite chiral states.}

From the viewpoint of chiral-mode control, the main conclusion is therefore that the { AEP organizes two related routes for chirality switching. The first is the direct AEP route, in which an infinitesimal phase-circulating perturbation switches the system from an achiral state to a chiral state through asymptotic chiral-mode selection. The second is the EP-pair route in the vicinity of the AEP, in which ordinary EPs enable switching between opposite chiral states.} These two routes are distinct in operation, but they are geometrically connected: as the perturbation amplitude required for the EP-pair route is reduced, the operating EPs approach the AEP along the exceptional lines, and the EP-pair route asymptotically approaches the same local exceptional-line geometry as the direct AEP route.

Within the present finite-resolution model, these two operating points exhibit different practical performance characteristics. In particular, under sufficiently high control resolution, the direct AEP operating point may provide a promising way for chirality generation with smaller perturbation amplitudes, and hence lower operating energy, than the EP-pair route. These results identify the AEP, rather than a generic DP or an ordinary EP, as the organizing structure from which multiple non-Hermitian routes to chiral-mode control can be designed.

\begin{acknowledgments}
This work was supported by the Japan Society for the Promotion of Science (Grants No. JP20H05641, No. JP21K14551, No. 24K01377, No. 24H02232, and No. 24H00400).
\end{acknowledgments}

\appendix

\section{Effective Hamiltonian for the three-resonator model}\label{app:A}

Here we derive the effective Hamiltonian used to analyze the chiral eigenmodes. We focus on the case $V=\varepsilon V_+$; the case $V=\varepsilon V_-$ is treated analogously.
\begin{gather}
  H(\varepsilon) = H_0 + \varepsilon V_+, \\
  \begin{aligned}
    \relax[H_0]_L = \diag(2, -1, -1), \qquad
    \relax[V_+]_L = \mqty(0 & 1 & 0 \\ 0 & 0 & 1 \\ 1 & 0 & 0).
  \end{aligned}
  \label{eq:2x2_perturbation}
\end{gather}
For the degenerate subspace $P=\ket{L_+}\bra{L_+}+\ket{L_-}\bra{L_-}$ and its complementary subspace $Q=I-P$, the corresponding $2\times2$ effective Hamiltonian is given by the Feshbach projection formalism as
\begin{align}
  H_{\mathrm{eff}}(\lambda)
  =
  PHP + PVQ \frac{1}{\lambda-QHQ} QVP .
  \label{eq:rigorous_Feshbach-Heff_for_app}
\end{align}
Here, $\lambda$ denotes an eigenvalue of the effective Hamiltonian.

The resolvent factor can be expanded perturbatively as a Neumann series,
\begin{align}
  H_{\mathrm{eff}}(\lambda)
  =
  PHP + PVQ R \sum_{n=0}^{\infty}(QVQR)^n QVP,
  \label{eq:schur_complement_neumann}
\end{align}
where $R := (\lambda-QHQ)^{-1}.$ Retaining only the lowest nontrivial order in the perturbation, we approximate the resolvent by its unperturbed value,
  $R \simeq R_0 := (\lambda_\pm(0)-QH_0Q)^{-1} = -{1}/3$, so that the effective Hamiltonian becomes
\begin{align}
  [H_{\mathrm{eff}}]_L
  \sim
  \mqty(
  -1                     & \varepsilon \\
  -\varepsilon^2/3 & -1
  ).
\end{align}
At leading order, the eigenvalues are therefore
\begin{align}
  \lambda_\pm(\varepsilon)
  = -1 
  \pm \frac{i}{\sqrt{3}} \varepsilon^{3/2}
  + O(\varepsilon^3),
\end{align}
which agrees with the result obtained directly from Eq.~\eqref{eq:3mode_eigenvalue_perturbation}. Thus, the two degenerate eigenvalues split according to the characteristic power law $\Delta\lambda \propto \varepsilon^{3/2}$. The corresponding eigenvectors $\ket{\psi_\pm}$ are, to leading order,
\begin{align}
  \ket{\psi_\pm}
  \propto
  \mqty(
  1 &&
  \mp \frac{i}{\sqrt{3}} \varepsilon^{1/2}
  )^\mathsf{T},
\end{align}
which shows explicitly that they asymptotically coalesce in the limit $\varepsilon \to 0$.

\section{N-site generalization and main results}
\label{APP:generalization_high_dimensional}

{ In this appendix we summarize the N-site extension and state the main results; the perturbative effective-Hamiltonian derivation is deferred to Appendix~\ref{app:eff}.}

\subsection{\texorpdfstring{$N$}{N}-site cyclic resonator system and degenerate OAM pairs}

We now generalize the preceding discussion to an $N$-dimensional tight-binding model with periodic boundary conditions. We consider a cyclic resonator array with symmetric couplings, where the coupling between resonators separated by $j$ sites is denoted by $\kappa_j$. In the site basis, the coupling matrix is
\begin{align}
  [K]_{\mathrm{site}}=
  \mqty(
  0        & \kappa_1 & \cdots   & \kappa_2 & \kappa_1 \\
  \kappa_1 & 0        & \kappa_1 &          & \kappa_2 \\
  \vdots   & \kappa_1 & 0        & \ddots   & \vdots   \\
  \kappa_2 &          & \ddots   & \ddots   & \kappa_1 \\
  \kappa_1 & \kappa_2 & \cdots   & \kappa_1 & 0
  ).
\end{align}
Let the complex onsite frequencies be $\omega_j \in \mathbb{C}$ $(j=0,\dots,N-1)$, and define
\[
  [\Omega]_{\mathrm{site}}=\mathrm{diag}(\omega_0,\omega_1,\dots,\omega_{N-1}).
\]
The model Hamiltonian is then given by
\[
  H_0=\Omega+K.
\]

When all onsite frequencies are identical, $\omega_j=\omega$ $(j=0,\dots,N-1)$, the Hamiltonian $H_0$ becomes circulant and can therefore be diagonalized by the discrete Fourier transform,
\begin{gather}
  U=\mqty(\ket{k_0} & \ket{k_1} & \cdots & \ket{k_{N-1}}), \\
  \ket{k_m}
  =
  \frac{1}{\sqrt{N}}
  \sum_{j=0}^{N-1}
  w_m^j \ket{j},
  \qquad
  w_m=e^{i2\pi m/N}.
\end{gather}
Here, $\ket{j}$ denotes the mode localized at site $j$. Because of periodicity, $\ket{k_m}=\ket{k_{m+N}}$. The corresponding eigenvalues are
\begin{align}
  \lambda_m
  =
  \omega+\sum_{j=0}^{N-1}\kappa_j w_m^j,
  \qquad
  m=0,\dots,N-1. \label{eq:high_dimensional_eigenvalues}
\end{align}

{{For a Hermitian cyclic array with symmetric couplings, $\kappa_j=\kappa_{N-j}$, the eigenvalues in Eq.~\eqref{eq:high_dimensional_eigenvalues} satisfy $$\lambda_{N-m} = \lambda_m,$$ where we used $w_{N-m}^{j}=w_m^{-j}=w_m^{N-j}$ and the symmetry $\kappa_j=\kappa_{N-j}$. 
Thus, the generic degeneracy structure is exhausted by the symmetry-related pair $(m,N-m)$.
Any higher-order degeneracy, involving additional indices beyond this pair, can occur only at special parameter values satisfying extra algebraic constraints among the couplings $\kappa_j$.
In the following, we restrict attention to the generic case in which each OAM doublet $(m,N-m)$ is spectrally isolated from all other modes.
}}
Accordingly, for even $N$, there are two nondegenerate eigenvalues at $m=0$ and $m=N/2$, whereas all others are doubly degenerate. For odd $N$, there is one nondegenerate eigenvalue at $m=0$, and all the others are doubly degenerate. These doubly degenerate mode pairs, $\ket{k_l}$ and $\ket{k_{N-l}}$, correspond to opposite OAM modes. We therefore define
% \[
\begin{align}
  \begin{aligned}
    \ket{L_l}:=\ket{k_l},
     &  &
    \ket{L_{-l}}:=\ket{k_{N-l}},
  \end{aligned}
\end{align}
% \]
with
% \[
  $l=1,\dots,l_{\mathrm{max}}$,  $l_{\mathrm{max}}= \lfloor{(N-1)/2}\rfloor$.
% \]
The chirality associated with the $l$th OAM pair is then defined as
\begin{align}
  \chi_j^{(l)}
  =
  \qty|\braket{L_l}{a_j}|-\qty|\braket{L_{-l}}{a_j}|,
\end{align}
where $\ket{a_j}$ denotes an eigenstate of the perturbed Hamiltonian. As in the previous sections, {{we measure all frequencies in units of an arbitrary positive reference coupling scale $\kappa_{\mathrm{ref}}>0$. Specifically, we define
\begin{align}
  \begin{aligned}
      \bar{H}_0
  :=
  \frac{H_0-\omega I}{\kappa_{\mathrm{ref}}},
  &&
  \bar{\kappa}_j
  :=
  \frac{\kappa_j}{\kappa_{\mathrm{ref}}},
  &&
  \bar{\lambda}_m
  :=
  \frac{\lambda_m-\omega}{\kappa_{\mathrm{ref}}}.
  \end{aligned}
\end{align}
In what follows, we omit the bars and use the same symbols for the dimensionless quantities. 
% The unperturbed Hamiltonian is then written as
% \begin{gather}
%   H_0
%   =
%   \sum_{m=0}^{N-1}\lambda_m \ket{k_m}\bra{k_m},\\
%   \begin{aligned}
%       \lambda_m
%   =
%   \sum_{j=0}^{N-1}\kappa_j w_m^j,
%   &&
%   m=0,\dots,N-1.
%   \end{aligned}
% \end{gather}}
}}

\subsection{AEP-based chirality generation in N-site systems}

The perturbations $V_\pm$ generalize to the $N$-site system as
\begin{align}
  \begin{gathered}
    V_+
    =
    \sum_{j=0}^{N-1}
    e^{-i2\pi j/N}\ket{j}\bra{j}
    =
    \sum_{j=0}^{N-1}
    \ket{k_{j-1}}\bra{k_j},
    \\
    V_-
    =
    \sum_{j=0}^{N-1}
    e^{i2\pi j/N}\ket{j}\bra{j}
    =
    \sum_{j=0}^{N-1}
    \ket{k_j}\bra{k_{j-1}}.
  \end{gathered}
\end{align}
Thus, in the angular-momentum ($k$-) space, $V_\pm$ act as unidirectional shift operators. The full Hamiltonian including both perturbations is
\begin{align}
  H
   & =
  H_0+\varepsilon_+V_++\varepsilon_-V_-
  \nonumber \\
   & =
  \sum_{j=0}^{N-1}
  \qty(
  \lambda_j \ket{k_j}\bra{k_j}
  +\varepsilon_+\ket{k_{j-1}}\bra{k_j}
  +\varepsilon_-\ket{k_j}\bra{k_{j-1}}
  ),
  \label{eq:high_dimensional_full_hamiltonian}
\end{align}
{{where $\varepsilon_\pm$ are dimensionless small parameters of the same order, measured in units of the reference coupling scale.}}

For the degenerate $l$th OAM pair, we define the projected subspace and its complement by
\begin{align}
  P
   & =
  \ket{L_l}\bra{L_l}+\ket{L_{-l}}\bra{L_{-l}},
  \qquad
  Q
  =
  I-P.
\end{align}
Using the effective-Hamiltonian construction summarized in Appendix \ref{app:eff}, we analyze the perturbative response of the Hermitian degeneracy to a unidirectional non-Hermitian perturbation. The effective Hamiltonian projected onto the $l$th OAM subspace is, to the lowest nontrivial order in $\varepsilon_+$,
\begin{align}
  [H_{\mathrm{eff}}^{(l)}]_L
  \simeq
  \mqty(
  \lambda_l &
  \varepsilon_+^{N-2l}
  \displaystyle\prod_{s=1}^{N-2l-1}
  \frac{1}{\lambda_l-\lambda_{l+s}}
  \\
  \varepsilon_+^{2l}
  \displaystyle\prod_{s=1}^{2l-1}
  \frac{1}{\lambda_l-\lambda_{l-s}}
            &
  \lambda_l
  ).
  \label{eq:high_dimensional_effective_hamiltonian}
\end{align}
Accordingly, the eigenvalue splitting and eigenvectors take the asymptotic forms
\begin{gather}
  \lambda_l^\pm-\lambda_l
  =
  \pm \sqrt{C_1C_2}\,\varepsilon_+^{N/2}
  +o\qty(\varepsilon_+^{N/2}),
  \label{eq:high_dimensional_eigenvalue_perturbation}
  \\
  [\ket*{\psi_\pm^{(l)}}]_L
  \sim
  \mqty(
  \sqrt{C_1}\,\varepsilon_+^{N/2-l} &
  \mp \sqrt{C_2}\,\varepsilon_+^{l}
  )^\mathsf{T}, \label{eq:high_dimensional_eigenvector_perturbation}
\end{gather}
where
\[
  C_1=
  \prod_{s=1}^{N-2l-1}\frac{1}{\lambda_l-\lambda_{l+s}},
  \qquad
  C_2=
  \prod_{s=1}^{2l-1}\frac{1}{\lambda_l-\lambda_{l-s}}.
\]
{{Here it is assumed that the $l$th OAM doublet is spectrally isolated, so that
$\lambda_l \neq \lambda_{l+s}$ for $s=1,\dots,N-2l-1$ and
$\lambda_l \neq \lambda_{l-s}$ for $s=1,\dots,2l-1$,
except for the symmetry-related equality $\lambda_l=\lambda_{N-l}$.
Under this non-resonance condition, the coefficients $C_1$ and $C_2$ are finite and the two-dimensional effective Hamiltonian is well defined.
If additional degeneracies are present, the above reduced description breaks down and one must instead perform degenerate perturbation theory in an enlarged subspace.}}

Equation~\eqref{eq:high_dimensional_eigenvalue_perturbation} shows that the eigenvalue response follows the fractional-power scaling $\Delta\lambda\sim \varepsilon_+^{N/2}$. As $N$ increases, the eigenvalue splitting becomes progressively higher order in the perturbation. 
{{For explicit numerical illustrations and fitting results, we now specialize to the nearest-neighbor model.}}
The fitted curves in Fig.~\ref{fig:ndim_powerplot} are of the form $y=ax^b$, and the fit parameters are summarized in Table~\ref{tab:fitting_parameters_high_dimensional}.

\begin{table}[tbhp]
  \raggedleft % 全体を右に寄せる

  % --- 上段：図のセクション ---
  \begin{minipage}{0.48\textwidth}
    \centering
    \includegraphics[height=4.5cm]{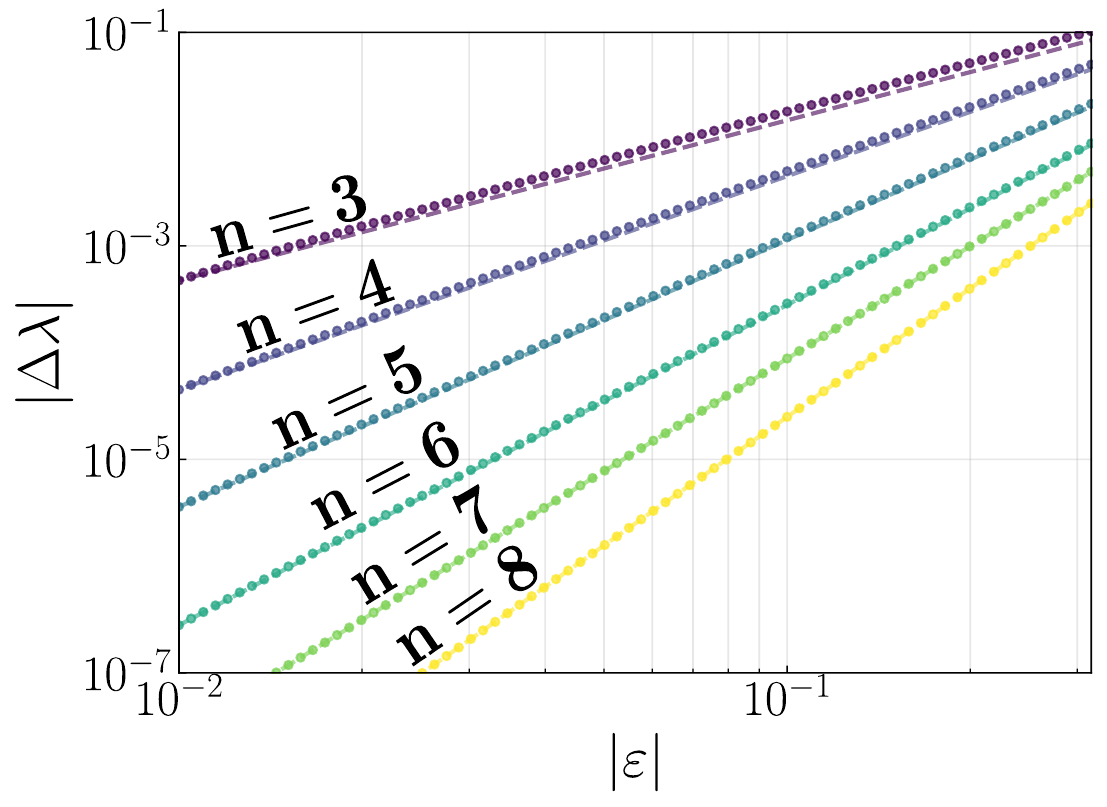}
    % \captionof を使うことで figure としてラベル付けする
    \captionof{figure}{Power-law dependence of the eigenvalue splitting on the perturbation strength {{for the nearest-neighbor model with}} $N=3,4,5,6,7,8$.}
    \label{fig:ndim_powerplot}
  \end{minipage}
  \vspace{1em} % 図と表の間の余白
  % --- 下段：表のセクション ---
  \begin{minipage}{0.48\textwidth}
    \centering
    \caption{Fitting parameters {{for the nearest-neighbor results shown}} in Fig.~\ref{fig:ndim_powerplot}, using $\Delta\lambda=A\varepsilon^p$. The results confirm $\Delta\lambda\sim\varepsilon^{N/2}$. The slight deviations for small $N$ arise from higher-order corrections of order $\varepsilon^N$, which are relatively more pronounced in that regime.
    }
    \label{tab:fitting_parameters_high_dimensional}
    \begin{ruledtabular}
      \begin{tabular}{ccc|ccc}
        $N$ & $A$     & $p$  & $N$ & $A$     & $p$  \\ \hline
        3   & 6.24e-1 & 1.54 & 6   & 2.92e-1 & 3.00 \\
        4   & 5.19e-1 & 2.02 & 7   & 2.79e-1 & 3.50 \\
        5   & 3.85e-1 & 2.51 & 8   & 2.50e-1 & 4.00 \\
      \end{tabular}
    \end{ruledtabular}
  \end{minipage}
\end{table}

The asymptotic form of the eigenvectors immediately yields the chiral-mode selection rule {{for the generic case $l\neq N/4$}}. If $N-2l>2l$, i.e., $l<N/4$, then the lower component dominates and
\begin{align}
  \lim_{\varepsilon_+\to0}[\ket*{\psi_\pm^{(l)}}]_L=(0,1)^\mathsf{T},
\end{align}
so the state {{asymptotically selects}} the negative-OAM mode. If $N-2l<2l$, i.e., $l>N/4$, then the upper component dominates and
\begin{align}
  \lim_{\varepsilon_+\to0}[\ket*{\psi_\pm^{(l)}}]_L=(1,0)^\mathsf{T},
\end{align}
so the state {{asymptotically selects}} the positive-OAM mode. {{Thus, for $l\neq N/4$, the sign of the selected chirality is determined solely by which of the two transition path lengths, $N-2l$ or $2l$, is shorter.}}

{{By contrast, when $N-2l=2l$, i.e., $l=N/4$, the two off-diagonal couplings in Eq.~\eqref{eq:high_dimensional_effective_hamiltonian} appear at the same perturbative order $O(\varepsilon_+^{N/2})$. In this borderline case, the order asymmetry responsible for the {{AEP}} chiral-mode selection disappears, and the leading-order eigenvectors do not asymptotically coalesce onto a single OAM eigenstate, but instead remain generic mixtures of the positive- and negative-OAM modes,
\begin{align}
  [\ket*{\psi_\pm^{(l)}}]_L
  \sim
  \mqty(
  \sqrt{C_1} &
  \pm \sqrt{C_2}
  )^\mathsf{T}.
\end{align} 
Therefore, for $l=N/4$, the {{AEP}} mechanism does not yield chiral-mode selection in general. 
For nearest-neighbor $N=4M$ rings, this borderline case is further special: an additional sublattice symmetry fixes the natural basis to achiral standing-wave combinations, so that $\chi^{(N/4)}=0$ holds exactly. A symmetry-based explanation of this exact achirality, together with its breakdown under symmetric beyond-nearest-neighbor couplings such as $\kappa_2$, is given in Appendix~\ref{app:achiral}.
}}

{{
Fig.~\ref{fig:OAM_Highern} compares the nearest-neighbor responses for $N=4$ and $N=5$ under the perturbation $V_+$. 
Figs.~\ref{fig:OAM_Highern}(a)--(c) correspond to the $N=4$ case, while Figs.~\ref{fig:OAM_Highern}(d)--(f) correspond to the $N=5$ case. 
For $N=4$, Fig.~\ref{fig:OAM_Highern}(a) shows the unperturbed spectrum, in which the degenerate pair $\ket{k_1},\ket{k_3}$ correspond to the borderline case $l=N/4=1$. 
Fig.~\ref{fig:OAM_Highern}(b) shows that the chirality of this pair remains zero as $\varepsilon_+\to0$, indicating the absence of asymptotic chiral-mode selection. 
This exact achirality is a special feature of the nearest-neighbor $N=4$ ring. As illustrated in Fig.~\ref{fig:OAM_Highern}(c), the two conversion paths between $\ket{k_1}$ and $\ket{k_3}$ have the same length, $2l=N-2l=2$, so the order asymmetry disappears, and the natural real-space basis is given by achiral standing-wave combinations.

By contrast, for $N=5$, Fig.~\ref{fig:OAM_Highern}(d) shows the unperturbed eigenvalues with two degenerate OAM pairs, $l=\pm1$ and $l=\pm2$. 
Figs.~\ref{fig:OAM_Highern}(e) and (f) show the corresponding chiralities $\chi^{(1)}$ and $\chi^{(2)}$, respectively. 
In this generic case $l\neq N/4$, strong chirality emerges under an infinitesimal perturbation, and the sign of the selected chirality is reversed between the two pairs, in agreement with the general selection rule: the asymptotically selected chirality changes sign depending on whether $l<N/4$ or $l>N/4$.}}
\begin{figure*}[tbph]
  \centering
  \includegraphics[width=\linewidth]{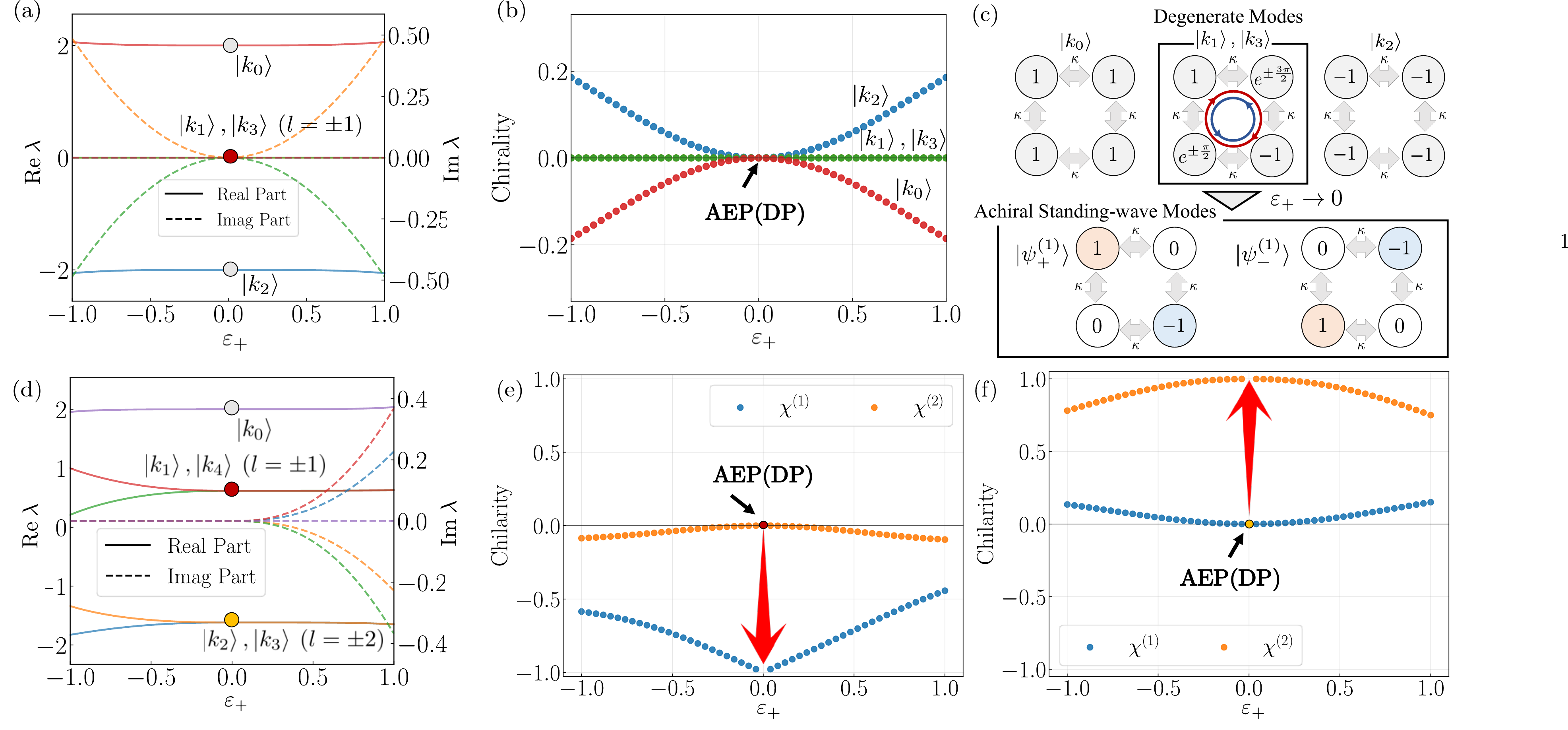}
  \caption{
Comparison between the $N=4$ borderline case and the generic $N=5$ case under the perturbation $V_+$. 
(a) Eigenvalues for $N=4$. The degenerate pair $\ket{k_1},\ket{k_3}$ corresponds to the special case $l=N/4=1$. 
(b) Chirality of the nearest-neighbor $N=4$ modes as a function of $\varepsilon_+$. The degenerate pair remains achiral in the limit $\varepsilon_+\to0$, showing the absence of asymptotic chiral-mode selection. 
(c) Real-space interpretation for the nearest-neighbor $N=4$ case: because the two conversion paths between $\ket{k_1}=\ket{L_{+1}}$ and $\ket{k_3}=\ket{L_{-1}}$ have the same length, the order asymmetry disappears; combined with the sublattice symmetry of the ring, this yields achiral standing-wave combinations rather than a single chiral OAM mode. 
(d) Eigenvalues for $N=5$, showing two degenerate OAM pairs, $\ket{k_1},\ket{k_4}$ $(l=\pm1)$ and $\ket{k_2},\ket{k_3}$ $(l=\pm2)$. 
(e) Chirality $\chi^{(1)}$ for the $l=\pm1$ pair in the $N=5$ case. 
(f) Chirality $\chi^{(2)}$ for the $l=\pm2$ pair in the $N=5$ case. 
For $N=5$, strong chirality emerges under an infinitesimal perturbation, and the selected chirality changes sign between the $l=\pm1$ and $l=\pm2$ pairs, consistent with the general rule for $l\neq N/4$.
}
  \label{fig:OAM_Highern}
\end{figure*}

\subsection{EP-based chirality reversal in N-site systems}

As discussed for the three-resonator system in Sec.~\ref{sec3:chiral_switching}, combining two perturbations enables chirality reversal. The $N$-site analogue of Eq.~\eqref{eq:3mode_hamiltonian_general_rotation} is obtained by parameterizing the perturbations in Eq.~\eqref{eq:high_dimensional_full_hamiltonian} as
\begin{align}
  \mqty(\varepsilon_+      \\ \varepsilon_-)
  =
  \mqty(
  \cos\theta  & \sin\theta \\
  -\sin\theta & \cos\theta
  )
  \mqty(\alpha             \\ \beta).
\end{align}
In particular, for $\theta=3\pi/4$, one has
\[
  \varepsilon_+=\frac{\beta'-\alpha'}{\sqrt{2}},
  \qquad
  \varepsilon_-=-\frac{\beta'+\alpha'}{\sqrt{2}},
\]
so that the onsite perturbation becomes
\begin{align}
  \mel{i}{V(3\pi/4)}{j}
  =
  -\sqrt{2}
  \qty(
  \alpha' \cos\frac{2\pi j}{N}
  +
  i\beta' \sin\frac{2\pi j}{N}
  )
  \delta_{ij},
\end{align}
which explicitly separates the real and imaginary onsite modulations.

Using Eq.~\eqref{eq:high_dimensional_effective_hamiltonian}, the off-diagonal matrix elements of the effective Hamiltonian in the $\{\ket{L_l},\ket{L_{-l}}\}$ basis are written as
\begin{align}
  \begin{aligned}
    A
     & =
    \mel{L_l}{H_{\mathrm{eff}}}{L_{-l}}
    \approx
    C_1 \varepsilon_+^{N-2l}
    +
    C_2 \varepsilon_-^{2l},
    \\
    B
     & =
    \mel{L_{-l}}{H_{\mathrm{eff}}}{L_l}
    \approx
    C_1 \varepsilon_-^{N-2l}
    +
    C_2 \varepsilon_+^{2l}.
  \end{aligned}
\end{align}
Let $m=\min(2l,N-2l)$. In the weak-perturbation regime, if $2l<N-2l$, then the leading-order contributions are
\begin{align}
  A
  \sim
  C_2\qty(\frac{\beta'+\alpha'}{\sqrt{2}})^{2l},
  \qquad
  B
  \sim
  C_2\qty(\frac{\beta'-\alpha'}{\sqrt{2}})^{2l}, \label{eq:high_dimensional_AB_2l}
\end{align}
whereas if $2l>N-2l$, they become
\begin{align}
  A
  \sim
  C_1\qty(\frac{\beta'-\alpha'}{\sqrt{2}})^{N-2l},
  \qquad
  B
  \sim
  C_1\qty(\frac{\beta'+\alpha'}{\sqrt{2}})^{N-2l}. \label{eq:high_dimensional_AB_N-2l}
\end{align}
Therefore, to leading order, the EP conditions reduce to $\alpha'=\beta'$ or $\alpha'=-\beta'$, corresponding to $A=0$ or $B=0$, respectively. The chirality associated with each EP follows the $(N,l)$-dependent chiral-selection rule derived above. Moreover, the real control amplitude required to move between EPs with opposite chirality is $\Delta\alpha' = 2\beta'$.
Thus, the chirality reversal induced by sweeping a purely real control parameter, found for $N=3$, persists in higher-dimensional systems.

{{Thus, in the weak-perturbation regime, the $N$-site system retains the same chirality-switching mechanism via {{ordinary}} EPs as in the three-resonator model, while the chirality selected at each EP is now governed by the $(N,l)$-dependent rule derived above.}}

\section{Derivation of the effective Hamiltonian in cyclic \texorpdfstring{$N$}{N}-site systems} \label{app:eff}

\begin{figure}[tbp]
  \centering
  \includegraphics[width=0.9\linewidth]{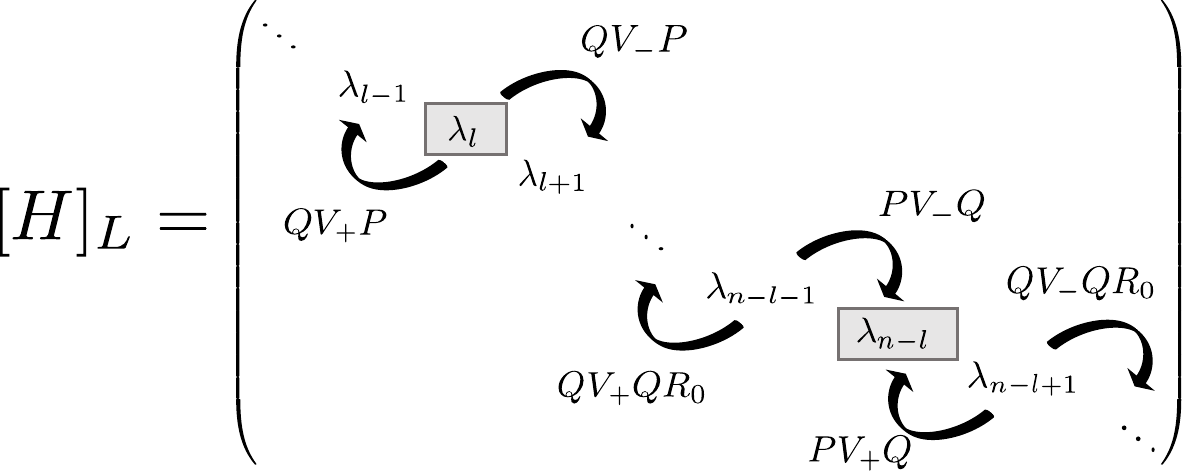}
  \caption{Transition processes between the target subspace $P$ and the complementary subspace $Q$ induced by the perturbations $V_\pm$. The minimum number of allowed steps required to connect $\ket{L_{\pm l}}$ and $\ket{L_{\mp l}}$ through the $Q$ space determines the perturbative order of the off-diagonal terms in the effective Hamiltonian.}
  \label{fig:QVQR0}
\end{figure}

{ This appendix derives Eqs.~\eqref{eq:high_dimensional_effective_hamiltonian}--\eqref{eq:high_dimensional_eigenvector_perturbation}.} The key point is that $V_+$ and $V_-$ act as unidirectional shift operators in angular-momentum space, so the leading off-diagonal matrix elements are determined by the shortest allowed transition paths connecting $\ket{L_l}$ and $\ket{L_{-l}}$ through the complementary subspace. As illustrated in Fig.~\ref{fig:QVQR0}, the two directions generally require different path lengths, which produces different perturbative orders for the two effective couplings. Starting from Eq.~\eqref{eq:high_dimensional_full_hamiltonian},
\begin{align}
  \begin{aligned}
    H
     & = \sum_{j=0}^{N-1}
    \qty[
      \lambda_j \ket{k_j}\bra{k_j}
      + \varepsilon_+ \ket{k_{j-1}}\bra{k_j}
      + \varepsilon_- \ket{k_j}\bra{k_{j-1}}
    ],
  \end{aligned}
\end{align}
For the degenerate subspace associated with the OAM pair of order $l$, we define $P = \ket{L_l}\bra{L_l} + \ket{L_{-l}}\bra{L_{-l}}, Q = I - P.$
The effective Hamiltonian is again constructed using Eq.~\eqref{eq:schur_complement_neumann}. The first-order projected perturbation is
\begin{align}
  PVP
  =
  \sum_{\sigma,\sigma'\in\{+,-\}}
  \ket{L_{\sigma l}}\bra{L_{\sigma l}}
  V
  \ket{L_{\sigma' l}}\bra{L_{\sigma' l}}.
\end{align}
Evaluating the matrix elements, one finds
\begin{gather*}
  \mel{L_l}{V}{L_l}=0,
  \qquad
  \mel{L_{-l}}{V}{L_{-l}}=0, \\
  \begin{aligned}
    \mel{L_l}{V}{L_{-l}}
     & =
    \varepsilon_+ \mel{k_l}{V_+}{k_{N-l}}
    + \varepsilon_- \mel{k_l}{V_-}{k_{N-l}} \\
     & =
    \varepsilon_+ \delta_{l,N-l-1}
    + \varepsilon_- \delta_{l,N-l+1}.
  \end{aligned}
\end{gather*}

For the second-order correction, because $H_0$ is diagonal in the $\ket{k_j}$ basis, the unperturbed resolvent is
\begin{align}
  R_0
  :=
  (\lambda_l - QH_0Q)^{-1}
  =
  \sum_{\substack{j=0 \\ j\neq l,\,N-l}}^{N-1}
  \frac{\ket{k_j}\bra{k_j}}{\lambda_l-\lambda_j}.
  \label{eq:R0}
\end{align}
% Introducing
% \begin{align}
%   S_Q =
%   \begin{cases}
%     \qty{0,1,\cdots,l_{\mathrm{max}},N/2}, & (N:\text{even}), \\
%     \qty{0,1,\cdots,l_{\mathrm{max}}},     & (N:\text{odd}),
%   \end{cases}
% \end{align}
Projecting onto the complementary subspace, we obtain
\begin{align}
  \begin{aligned}
    PVQ
     & =
    \sum_{\sigma\in\{+,-\}}
    \sum_{\substack{j=0 \\ j\neq l,\,N-l}}^{N-1}
    \ket{L_{\sigma l}}\bra{L_{\sigma l}} V \ket{k_j}\bra{k_j}\\
     & =
    \varepsilon_+\ket{L_l}\bra{L_{l+1}}
    + \varepsilon_-\ket{L_l}\bra{L_{l-1}}       \\
     & \quad
    + \varepsilon_+\ket{L_{-l}}\bra{L_{-l+1}}
    + \varepsilon_-\ket{L_{-l}}\bra{L_{-l-1}}.
  \end{aligned}
  \label{eq:PVQ}
\end{align}
This term describes transitions from neighboring sectors back into the target OAM pair.

Similarly,
\begin{align}
  \begin{aligned}
    QVP
     & =
    \sum_{\sigma\in\{+,-\}}
    \sum_{\substack{j=0 \\ j\neq l,\,N-l}}^{N-1}
    \ket{k_j}\bra{k_j} V \ket{L_{\sigma l}}\bra{L_{\sigma l}}\\
     & =
    \varepsilon_+\ket{L_{l-1}}\bra{L_l}
    + \varepsilon_-\ket{L_{l+1}}\bra{L_l}       \\
     & \quad
    + \varepsilon_+\ket{L_{-l-1}}\bra{L_{-l}}
    + \varepsilon_-\ket{L_{-l+1}}\bra{L_{-l}}.
  \end{aligned}
  \label{eq:QVP}
\end{align}
This term describes transitions from the target OAM pair to neighboring sectors.

Next, defining $S_\pm = \qty{l,\,N-l,\,l\pm1,\,N-l\pm1}$, we have
\begin{align}
  QVQ R_0
  =
  \varepsilon_+ \sum_{j\notin S_+}^{N-1}
  \frac{\ket{k_{j-1}}\bra{k_j}}{\lambda_l-\lambda_j}
  +
  \varepsilon_- \sum_{j\notin S_-}^{N-1}
  \frac{\ket{k_{j+1}}\bra{k_j}}{\lambda_l-\lambda_j}.
  \label{eq:QVQR0}
\end{align}

The structure of these terms admits a simple interpretation, illustrated in Fig.~\ref{fig:QVQR0}. The operator $QVP$ corresponds to the first step, taking the system from $\ket{k_l}$ or $\ket{k_{N-l}}$ to neighboring states such as $\ket{k_{l\pm1}}$ or $\ket{k_{N-l\pm1}}$, with one power of the perturbation. The factor $(QVQ R_0)^p$ then describes $p$ additional allowed steps within the $Q$ subspace, each accompanied by both a perturbation factor and an energy denominator. Finally, $PVQ$ returns the system from $\ket{k_{l\pm1}}$ or $\ket{k_{N-l\pm1}}$ to the target subspace $P$. Therefore, the leading perturbative order contributing to an off-diagonal element of the effective Hamiltonian is determined by the minimum number of allowed steps required to connect $\ket{L_{\pm l}}$ and $\ket{L_{\mp l}}$ through the $Q$ subspace.

Since $V_+$ and $V_-$ generate distinct transition processes and do not mix at the lowest order, we first consider them separately. With only $V_+$ present, the transition
\[
  \ket{L_l}=\ket{k_l} \to \ket{L_{-l}}=\ket{k_{N-l}}
\]
proceeds along the path $l \to l-1 \to \cdots \to N-l+1 \to N-l$. The number of intermediate transitions within $Q$ is therefore $N-2l-2$, so the total perturbative order is $N-2l$. Conversely, the reverse transition
\[
  \ket{L_{-l}}=\ket{k_{N-l}} \to \ket{L_l}=\ket{k_l}
\]
proceeds along $N-l \to N-l-1 \to \cdots \to l+1 \to l$, which contains $2l-2$ intermediate $Q$-space steps and hence appears at order $2l$.

Therefore, using only $V_+$, the relevant off-diagonal elements of the effective Hamiltonian are
\begin{align}
  \begin{aligned}
    \mel{L_l}{H_{\mathrm{eff}}}{L_{-l}}
     & =
    \varepsilon_+^{N-2l}
    \prod_{s=1}^{N-2l-1}
    \frac{1}{\lambda_l-\lambda_{l+s}}, \\
    \mel{L_{-l}}{H_{\mathrm{eff}}}{L_l}
     & =
    \varepsilon_+^{2l}
    \prod_{s=1}^{2l-1}
    \frac{1}{\lambda_l-\lambda_{l-s}}.
  \end{aligned}
\end{align}
Similarly, using only $V_-$, one obtains
\begin{align}
  \begin{aligned}
    \mel{L_l}{H_{\mathrm{eff}}}{L_{-l}}
     & =
    \varepsilon_-^{2l}
    \prod_{s=1}^{2l-1}
    \frac{1}{\lambda_l-\lambda_{l-s}}, \\
    \mel{L_{-l}}{H_{\mathrm{eff}}}{L_l}
     & =
    \varepsilon_-^{N-2l}
    \prod_{s=1}^{N-2l-1}
    \frac{1}{\lambda_l-\lambda_{l+s}}.
  \end{aligned}
\end{align}
Accordingly, when both $V_+$ and $V_-$ are present, the leading perturbative order is governed by $\min(2l,N-2l)$, and the off-diagonal elements behave as
\begin{align}
  \begin{aligned}
    \mel{L_l}{H_{\mathrm{eff}}}{L_{-l}}
     & \sim
    \begin{cases}
      \varepsilon_-^{2l},                      & (2l < N-2l), \\
      \varepsilon_+^{N-2l},                        & (2l > N-2l), \\
      \varepsilon_+^{N/2} + \varepsilon_-^{N/2}, & (2l = N-2l),
    \end{cases}
    \\
    \mel{L_{-l}}{H_{\mathrm{eff}}}{L_l}
     & \sim
    \begin{cases}
      \varepsilon_+^{2l},                      & (2l < N-2l), \\
      \varepsilon_-^{N-2l},                        & (2l > N-2l), \\
      \varepsilon_+^{N/2} + \varepsilon_-^{N/2}, & (2l = N-2l).
    \end{cases}
  \end{aligned}
  \label{eq:high_dimensional_off_diagonal_correction}
\end{align}

The diagonal terms first appear at second order. Using Eqs.~\eqref{eq:R0}, \eqref{eq:PVQ}, and \eqref{eq:QVP}, one finds
\begin{align}
  \begin{aligned}
    &\mel{L_{\pm l}}{H_{\mathrm{eff}}}{L_{\pm l}} -   \lambda_l\\
     \quad & =
    \varepsilon_+\varepsilon_-
    \qty(
    \frac{1}{\lambda_l-\lambda_{l-1}}
    +
    \frac{1}{\lambda_l-\lambda_{l+1}}
    )
    + o(\varepsilon_+\varepsilon_-).
  \end{aligned}
  \label{eq:high_dimensional_diagonal_correction}
\end{align}
Thus, the diagonal correction appears only at second order. Physically, this corresponds to a process in which the state leaves the target subspace $P$, enters $Q$, and immediately returns. Such a process requires both $V_+$ and $V_-$: one lowers the angular-momentum label and the other raises it. Hence, no diagonal correction appears at any order if only one of the two perturbations is present.

The eigenvalues of the effective Hamiltonian are therefore given, at the lowest order, by
\begin{align}
  \lambda_\pm^{(l)}
  =
  \mel{L_l}{H_{\mathrm{eff}}}{L_l}
  \pm
  \sqrt{
    \mel{L_l}{H_{\mathrm{eff}}}{L_{-l}}
    \mel{L_{-l}}{H_{\mathrm{eff}}}{L_l}
  }.
\end{align}
Accordingly, the splitting behaves as
\begin{align}
  \begin{aligned}
    \sqrt{D_l}
     & :=
    \sqrt{
      \mel{L_l}{H_{\mathrm{eff}}}{L_{-l}}
      \mel{L_{-l}}{H_{\mathrm{eff}}}{L_l}
    }
    \\
     & \sim
    \begin{cases}
      (\varepsilon_+\varepsilon_-)^{N/2-l},    & (l < N/4), \\
      (\varepsilon_+\varepsilon_-)^l,          & (l > N/4), \\
      \varepsilon_+^{N/2}+\varepsilon_-^{N/2}, & (l = N/4).
    \end{cases}
  \end{aligned}
  \label{eq:high_dimensional_eigenvalue_splitting}
\end{align}

Finally, let us return to the case $\varepsilon_-=0$ in order to examine the perturbative response of higher-order {{chiral modes}}. The effective Hamiltonian projected onto the degenerate OAM subspace of order $l$ then becomes, to the lowest nontrivial order in $\varepsilon_+$,
\begin{align}
  [H_{\mathrm{eff}}^{(l)}]_L
  \simeq
  \mqty(
  \lambda_l &
  \varepsilon_+^{N-2l}\displaystyle\prod_{s=1}^{N-2l-1}\frac{1}{\lambda_l-\lambda_{l+s}}
  \\
  \varepsilon_+^{2l}\displaystyle\prod_{s=1}^{2l-1}\frac{1}{\lambda_l-\lambda_{l-s}}
            &
  \lambda_l
  ).
\end{align}

\section{Symmetry-protected achirality of the nearest-neighbor $N=4M$ case}
\label{app:achiral}

For the borderline pair $l=N/4$, Eq.~\eqref{eq:high_dimensional_effective_hamiltonian}
already shows that the two opposite conversion processes appear at the same perturbative order
$O(\varepsilon^{N/2})$. Therefore, the order asymmetry responsible for asymptotic chiral
selection disappears, and the limiting states remain noncoalesced mixtures in general.

The nearest-neighbor $N=4M$ ring is a further special case.
For
\begin{align}
  H_0-\omega I
  =
  \kappa_1
  \sum_{j=0}^{N-1}
  \qty(\ket{j}\bra{j+1}+\ket{j+1}\bra{j}),
\end{align}
the unperturbed eigenvalues are $\lambda_m = 2\kappa_1 \cos\qty({2\pi m}/{N})$. At the borderline index $l=N/4$, one has $\lambda_{N/4}=0$, and for
$s=1,\dots,N/2-1$,
% \begin{align}
%   \lambda_{N/4+s}
%   &=
%   2\kappa_1 \cos\qty(\frac{\pi}{2}+\frac{2\pi s}{N})
%   =
%   -2\kappa_1 \sin\qty(\frac{2\pi s}{N}),
%   \\
%   \lambda_{N/4-s}
%   &=
%   2\kappa_1 \cos\qty(\frac{\pi}{2}-\frac{2\pi s}{N})
%   =
%   +2\kappa_1 \sin\qty(\frac{2\pi s}{N}).
% \end{align}
\begin{align}
  \lambda_{N/4\pm s}
  &=
  2\kappa_1 \cos\qty(\frac{\pi}{2}\pm\frac{2\pi s}{N})
  =
  \mp 2\kappa_1 \sin\qty(\frac{2\pi s}{N}).
\end{align}
Hence $\lambda_{N/4+s} = -\lambda_{N/4-s}.$ Substituting this into the coefficients in
Eq.~\eqref{eq:high_dimensional_effective_hamiltonian}, one finds
\begin{align}
  C_1
  &=
  \prod_{s=1}^{N/2-1}
  \frac{1}{\lambda_{N/4}-\lambda_{N/4+s}},
  \\
  C_2
  &=
  \prod_{s=1}^{N/2-1}
  \frac{1}{\lambda_{N/4}-\lambda_{N/4-s}}
  =
  (-1)^{N/2-1} C_1.
\end{align}
Since $N=4M$ implies $N/2-1=2M-1$ is odd, this reduces to $C_2 = - C_1, |C_1|=|C_2|$.

Therefore the leading-order eigenvectors in the $\{\ket{L_{N/4}},\ket{L_{-N/4}}\}$ basis
have equal-weight components,
\begin{align}
  [\ket*{\psi_\pm^{(N/4)}}]_L
  \sim
  \mqty(
  \sqrt{C_1} &
  \pm \sqrt{C_2}
  )^\mathsf{T},
\end{align}
and can always be written, after an appropriate phase choice within the degenerate doublet,
as equal-weight superpositions of $\ket{L_{N/4}}$ and $\ket{L_{-N/4}}$.
Thus, in the nearest-neighbor $N=4M$ case, the borderline pair is asymptotically achiral in the limit
$\varepsilon_+\to0$.

This result admits a simple real-space interpretation.
Define the sublattice operator
\begin{align}
  \Gamma
  :=
  \sum_{j=0}^{N-1}
  (-1)^j \ket{j}\bra{j}.
\end{align}
For the nearest-neighbor ring it satisfies
\begin{align}
  \Gamma^2=I,
  \qquad
  \qty{\Gamma,\,H_0-\omega I}=0.
\end{align}
Moreover, in the angular-momentum basis one has $\Gamma\ket{k_m}=\ket{k_{m+N/2}}$, so that, for $l=N/4$,
\begin{align}
  \begin{aligned}
      \Gamma\ket{L_{N/4}}=\ket{L_{-N/4}},
  &&
  \Gamma\ket{L_{-N/4}}=\ket{L_{N/4}}.
  \end{aligned}
\end{align}

The associated standing-wave combinations are
\begin{align}
  \begin{aligned}
    \ket{C}
    &:=
    \frac{\ket{L_{N/4}}+\ket{L_{-N/4}}}{\sqrt2}, &&
    \ket{S}
    &:=
    \frac{\ket{L_{N/4}}-\ket{L_{-N/4}}}{i\sqrt2},
  \end{aligned}
\end{align}
which satisfy $\Gamma\ket{C}=+\ket{C}$, $\Gamma\ket{S}=-\ket{S}$.
On the other hand, the lowest-order process that returns the state to the $l=N/4$ subspace
is generated after $N/2$ applications of $V_+$, giving
\begin{align}
  V_+^{N/2}
  =
  \sum_{j=0}^{N-1}
  e^{-i\pi j}\ket{j}\bra{j}
  =
  \sum_{j=0}^{N-1}
  (-1)^j\ket{j}\bra{j}
  =
  \Gamma.
\end{align}
Thus the lowest-order perturbation returning to the borderline doublet is compatible with the
standing-wave basis selected by $\Gamma$, rather than with a single chiral OAM mode.

Once a symmetric beyond-nearest-neighbor coupling connecting the same sublattice is introduced,
this special structure is generally lost.
For example, a next-nearest-neighbor coupling $\kappa_2$ breaks the anticommutation relation
$\qty{\Gamma,H_0-\omega I}=0$ and, more importantly, destroys the spectral antisymmetry
$\lambda_{N/4+s}=-\lambda_{N/4-s}$ responsible for $|C_1|=|C_2|$.
Then the two opposite processes still appear at the same order $O(\varepsilon_+^{N/2})$,
so asymptotic chiral-mode selection does not reappear, but the limiting states need no longer be
equal-weight superpositions.
Accordingly, the borderline pair remains noncoalesced, while its chirality is generically nonzero.

\section{Finite-resolution averaging and explicit FoM definitions}
\label{app:avg_fom}

In the main text, we discussed the operating points associated with the two chirality-switching principles { organized by the AEP} under finite control resolution. Since the ideal AEP-based achiral-to-chiral switching response is defined in the infinitesimal-perturbation limit, it is important to clarify how the observed chirality depends on the averaging scheme and how the figure of merit (FoM) should be defined explicitly for the { direct AEP and EP-pair operating points}. In this Appendix, we introduce several representative averaging schemes and show that the qualitative conclusions presented in the main text are robust with respect to these choices.

\subsection{Explicit definitions of \texorpdfstring{$\mathrm{FoM}_{\mathrm{AEP}}$}{FoM\_AEP} and \texorpdfstring{$\mathrm{FoM}_{\mathrm{EP}}$}{FoM\_EP}}
\label{app:fom_explicit}

To compare the { direct AEP and EP-pair operating points} as chirality switching, we introduced in the main text the engineering figure of merit
\begin{align}
  \mathrm{FoM}
  =
  \frac{C(\rho)}
  {(1+\Delta(\rho))(1+\Gamma(\rho))}
  \label{eq:fom_general_appendix}
\end{align}
Here, $C(\rho)$ denotes the normalized chirality-change ratio defined in Eq.~\eqref{eq:chirality_change_ratio}, $\Delta(\rho)$ is the required sweep amplitude along the real-part control direction, and $\Gamma(\rho)$ represents the corresponding cost in the imaginary-part direction, namely, the required gain/loss component.

For the { AEP} operating point, the perturbation $V_\mathrm{AEP}$ can be decomposed into its real and imaginary parts as
\begin{align}
  V_\mathrm{AEP}(\alpha)
   & = \alpha V_+ = 
  - \frac{\alpha}{\sqrt{2}} V_\mathrm{R}
  +
  i \frac{\alpha}{\sqrt{2}}V_\mathrm{I}
  \label{eq:Vplus_real_imag}
\end{align}
where $V_\mathrm{R}$ and $V_\mathrm{I}$ are defined as
% \begin{align}
%   [V_\mathrm{R}]_\mathrm{site}
%   &:=
%   \begin{pmatrix}
%     -\sqrt{2} & 0                  & 0                  \\
%     0         & \frac{1}{\sqrt{2}} & 0                  \\
%     0         & 0                  & \frac{1}{\sqrt{2}}
%   \end{pmatrix},\\
%   [V_\mathrm{I}]_\mathrm{site}
%   &:=
%   \begin{pmatrix}
%     0 & 0                   & 0                  \\
%     0 & -\sqrt{\frac{3}{2}} & 0                  \\
%     0 & 0                   & \sqrt{\frac{3}{2}}
%   \end{pmatrix}.
% \end{align}
\begin{align}
  [V_\mathrm{R}]_\mathrm{site}
  &:=
  \diag(-\sqrt{2}, \:  {1}/{\sqrt{2}}, \: {1}/{\sqrt{2}}),
  \\
  [V_\mathrm{I}]_\mathrm{site}
  &:=
  \diag(0, \: -\sqrt{3/2}, \: \sqrt{3/2}).
\end{align}
This decomposition shows that, at the { AEP} operating point, the real and imaginary perturbation components vary simultaneously along the $V_+$ direction.

By contrast, in the EP-based switching protocol,
\begin{align}
  V_\mathrm{EP}(\alpha',\beta')
   & = V(3\pi/4) = 
  \alpha'V_\mathrm{R}
  +
  \beta'V_\mathrm{I}
  \label{eq:V_alpha_beta_appendix}
\end{align}
so that the real and imaginary control parameters can be tuned independently.

Using the quantities $C_{\mathrm{AEP}}(\rho)$, $C_{\mathrm{EP}}(\rho)$, $\Delta\alpha_{\mathrm{AEP}}(\rho)$, and $\Delta\alpha'_{\mathrm{EP}}(\rho)$ extracted from Fig.~\ref{fig:rhoboke}(b), (c) and Eq.~\eqref{eq:chirality_change_ratio}, we define
\begin{align}
  \mathrm{FoM}_{\mathrm{AEP}} (\rho)
   & =
  \frac{
  C_{\mathrm{AEP}}
  }{
  \left(1+\Delta\alpha_{\mathrm{AEP}}/\sqrt{2}\right)
  \left(1+\Delta\alpha_{\mathrm{AEP}}/\sqrt{2}\right)
  }, \notag
  \\
  \mathrm{FoM}_{\mathrm{EP}} (\rho)
   & =
  \frac{
  C_{\mathrm{EP}}
  }{
  \left(1+\Delta\alpha'_{\mathrm{EP}}\right)
  \left(1+\beta'\right)
  }
  \label{eq:fom_ep_explicit}
\end{align}
The first factor in the denominator reflects the required control amplitude along the real direction, while the second factor accounts for the cost of implementing gain/loss together with the effect of finite uncertainty width. Under this definition, the FoM of the { direct AEP} operating point can exceed that of the { EP-pair} operating point in the high-resolution limit, whereas the EP operating point becomes more favorable when the available control precision is insufficient. This comparison supports the view that the {{AEP}}-based operating principle can outperform EP-based control under highly accurate parameter tuning. { The EP-based FoM defined above corresponds to the full chirality-reversal process used in the main text. In the following subsection, we introduce a complementary single-branch perturbation-sweep definition, in which the same EP branch is evaluated as an achiral-to-chiral operation.}

{ \subsection{Complementary achiral-to-chiral FoM for the EP branch}\label{app:alternatives_gene}}

For the main-text comparison, the EP-based operating point is treated in its natural role as a full chirality-reversal process between two opposite-chirality states. As a direct comparison with the achiral-to-chiral operation via the AEP, we also consider a complementary single-branch perturbation-sweep interpretation of the same EP branch. In this interpretation, the EP response is evaluated as a chirality-generation process from a chirality-neutral reference point to one EP-associated chiral state.

For a fixed loss bias $\beta'$, let $\alpha_0'$ denote the chirality-neutral reference point of the resolution-averaged response, defined by
\begin{align}
  \ev{\chi}_{\mathrm{disk}}(\alpha_0',\beta')=0 .
  \label{eq:alpha0_definition}
\end{align}
This point is used as the reference state for the EP-based achiral-to-chiral operation.

\begin{figure}[tbp]
  \centering
  \includegraphics[width=0.9\linewidth]{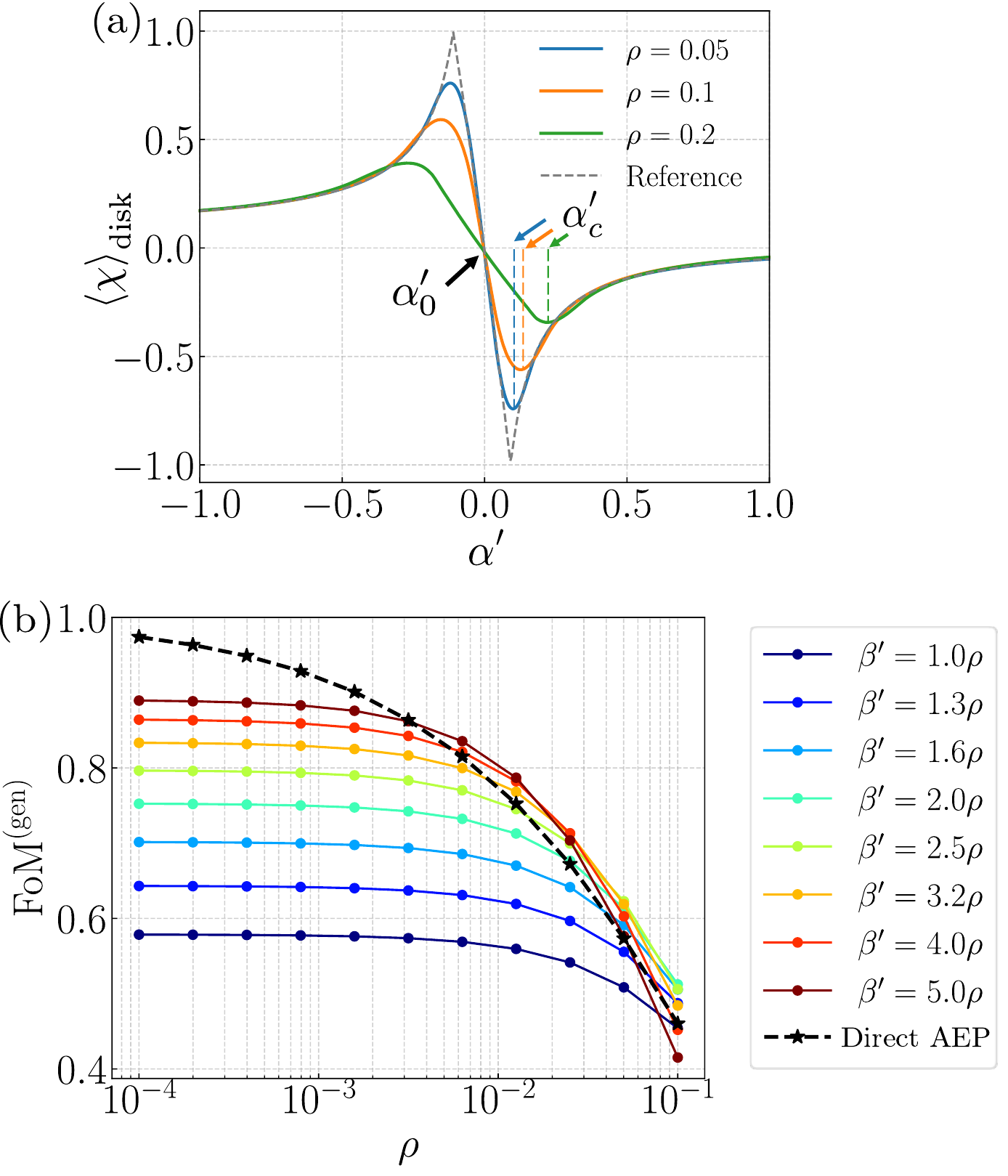}
  \caption{
  Complementary single-branch achiral-to-chiral FoM for the EP branch.
  (a) Disk-averaged chirality $\ev{\chi}_{\mathrm{disk}}$ as a function of $\alpha'$ for a fixed $\beta'$. The red arrows indicate the chirality-generation contrast $\Delta \ev{\chi}^{\mathrm{gen}}_{\mathrm{EP}}$ and the corresponding real-control excursion $\Delta\alpha_{\mathrm{EP}}^{\prime,\mathrm{gen}}$, measured from the chirality-neutral reference point $\alpha'_0$ to the negative-chirality EP-associated extremum $\alpha'_{c,-}$.
  (b) Figure of merit for the achiral-to-chiral comparison. The AEP-based result is compared with the single-branch EP-based result defined in Eqs.~\eqref{eq:delta_chi_ep_gen}--\eqref{eq:fom_ep_gen}. Within the present averaging and cost model, the AEP-based route can become favorable in the sufficiently high-resolution regime.
  }
  \label{fig:fom_ep_alternative}
\end{figure}

We then define $\alpha_{c,-}'$ as the EP-associated extremum on the negative-chirality branch,
\begin{align}
  \ev{\chi}_{\mathrm{disk}}(\alpha_{c,-}',\beta')<0 .
\end{align}
In the numerical evaluation shown in Fig.~\ref{fig:fom_ep_alternative}, we use this negative-chirality branch because, in the finite parameter range considered here, it gives a slightly more favorable single-branch response from the chirality-neutral reference point. This choice does not affect the high-resolution conclusion: as $\rho$ becomes small and the optimal $\beta'$ approaches zero, the positive- and negative-chirality branches become nearly symmetric.

The single-branch chirality change is defined as
\begin{align}
  \Delta\ev{\chi}_{\mathrm{EP}}^{\mathrm{gen}}
   &:=
  \abs{
  \ev{\chi}_{\mathrm{disk}}(\alpha_{c,-}',\beta')
  -
  \ev{\chi}_{\mathrm{disk}}(\alpha_0',\beta')
  } ,
  \label{eq:delta_chi_ep_gen}
\end{align}
and the corresponding real-control excursion is defined as
\begin{align}
  \Delta\alpha_{\mathrm{EP}}^{\prime,\mathrm{gen}}
   &:=
  \abs{\alpha_{c,-}'-\alpha_0'} .
  \label{eq:delta_alpha_ep_gen}
\end{align}
Since this single-branch process represents switching from a chirality-neutral state to a chiral state, its ideal chirality change is unity. We therefore define the normalized chirality-change ratio as $C_{\mathrm{EP}}^{\mathrm{gen}}(\rho)=  \Delta\ev{\chi}_{\mathrm{EP}}^{\mathrm{gen}}$. The corresponding figure of merit is then
\begin{align}
  \mathrm{FoM}_{\mathrm{EP}}^{\mathrm{gen}}(\rho)
   &=
  \frac{
  C_{\mathrm{EP}}^{\mathrm{gen}}(\rho)
  }{
  \left(1+\Delta\alpha_{\mathrm{EP}}^{\prime,\mathrm{gen}}(\rho)\right)
  \left(1+\beta'\right)
  } .
  \label{eq:fom_ep_gen}
\end{align}
Here, the first factor in the denominator represents the real-part control excursion required to reach one EP-associated chiral state from the chirality-neutral point, while the second factor accounts for the finite imaginary bias.

This chirality-generation metric differs from the full reversal metric used in the main text only in the definition of the EP operating range. When the averaged chirality curve is approximately antisymmetric about the midpoint between the two EP-associated extrema, one has
\begin{align}
  \begin{aligned}
    \Delta\ev{\chi}_{\mathrm{EP}}^{\mathrm{rev}}
    \simeq
    2\Delta\ev{\chi}_{\mathrm{EP}}^{\mathrm{gen}}, &&
    \Delta\alpha_{\mathrm{EP}}^{\prime,\mathrm{rev}}
    \simeq
    2\Delta\alpha_{\mathrm{EP}}^{\prime,\mathrm{gen}} ,
  \end{aligned}
  \label{eq:relation_gen_rev}
\end{align}
where the superscript $\mathrm{rev}$ denotes the full chirality-reversal quantities used in the main text.

Figure~\ref{fig:fom_ep_alternative}(a) illustrates the geometric definitions of $\Delta\ev{\chi}_{\mathrm{EP}}^{\mathrm{gen}}$ and $\Delta\alpha_{\mathrm{EP}}^{\prime,\mathrm{gen}}$ on the resolution-averaged chirality curve. Figure~\ref{fig:fom_ep_alternative}(b) shows the corresponding figure of merit. Although this single-branch definition changes the quantitative balance between the AEP and EP operating points, it does not alter the qualitative conclusion of the present work: within the present averaging and cost model, the AEP-based achiral-to-chiral route can become favorable in the sufficiently high-resolution regime.

\subsection{One-dimensional averaging along the loss direction}
\label{app:avg_loss_only}

{ We next examine the robustness of the above definitions with respect to the averaging scheme by considering one-dimensional broadening only along the imaginary-direction control axis.}

As the simplest model for the case in which gain/loss uncertainty is the dominant source of control error, we consider averaging only along the imaginary-direction control parameter. Let $(\alpha',\beta')$ denote the real and imaginary perturbation amplitudes defined in Eq.~\eqref{eq:perturbation_135_local}. The averaged chirality is then defined by
\begin{align}
  \expval{\chi}_{\mathrm{imag}}(\alpha',\beta')
  =
  \frac{1}{2\rho}
  \int_{-\rho}^{\rho}
  \chi(\alpha',\beta' + \Delta\beta')\, d\Delta\beta'.
\end{align}
Here, $\rho$ is the half-width of the uncertainty along the loss direction and is normalized by the coupling {{rate}}.

This one-dimensional averaging describes a situation in which the real-part control is sufficiently well calibrated, whereas the dominant uncertainty arises from the implementation of loss or gain. Even under this averaging scheme, the qualitative behavior discussed in the main text remains unchanged. Specifically, the { AEP} operating point is smoothed into an effectively chirality-neutral point because the positive- and negative-chirality regions are mixed by the averaging. 

% Because this averaging changes the definition of $\expval{\chi}_{\rho}$, accordingly, slightly modified as
% \begin{align}
%   \mathrm{FoM}_{\mathrm{AEP}}^{\mathrm{(imag)}}
%    & =
%   \frac{
%   C_{\mathrm{AEP}}(\rho)
%   }{
%   \left(1+\Delta\alpha_{\mathrm{AEP}}/\sqrt{2}\right)
%   \left(1+\Delta\alpha_{\mathrm{AEP}}/\sqrt{2}+\rho\right)
%   }, \notag
%   \\
%   \mathrm{FoM}_{\mathrm{EP}}^{\mathrm{(imag)}}
%    & =
%   \frac{
%   C_{\mathrm{EP}}(\rho)
%   }{
%   \left(1+\Delta\alpha'_{\mathrm{EP}}\right)
%   \left(1+\beta' + \rho\right)
%   }
%   \label{eq:fom_ep_explicit_2}
% \end{align}

In this case as well, the same device-oriented metric can be applied to the present one-dimensional averaging scheme. What changes is not the structure of the FoM, but the resolution-averaged chirality profile from which the quantities entering it are extracted. In particular, the values of $\Delta\ev{\chi}_{\mathrm{AEP}}$, $\Delta\ev{\chi}_{\mathrm{EP}}$, $\Delta\alpha_{\mathrm{AEP}}$, and $\Delta\alpha'_{\mathrm{EP}}$ are evaluated using $\ev{\chi}_{\mathrm{imag}}$ instead of $\ev{\chi}_{\mathrm{disk}}$.

\begin{figure}[tbph]
  \centering
  \includegraphics[height=4.5cm]{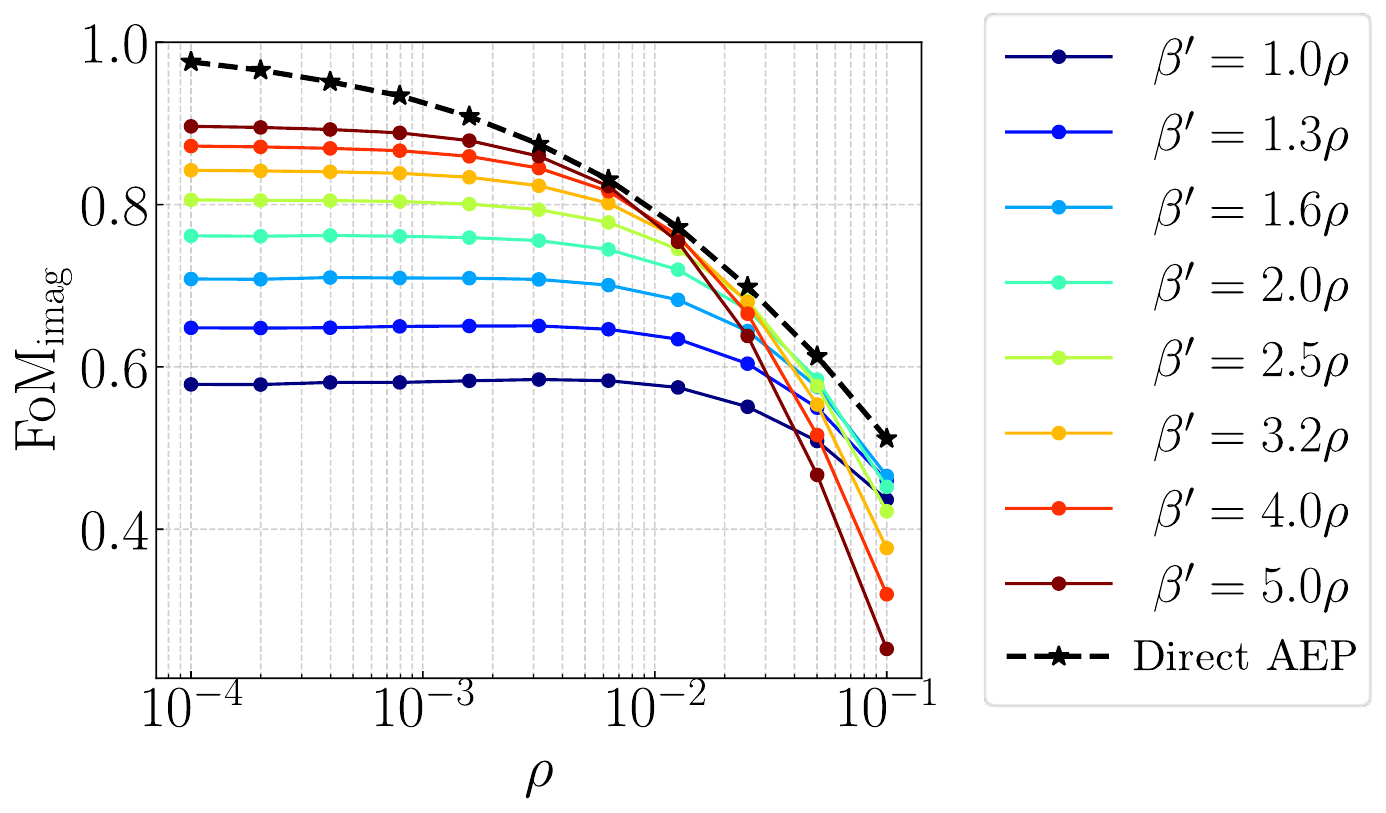}
  \caption{Dependence of the figure of merit (FoM) on $\rho$, where the broadening is applied along the loss direction. The result is qualitatively consistent with that in the main text (Fig.~\ref{fig:rhoboke}), demonstrating that the distinction between { the direct AEP and EP-pair
operating points is robust against the specific choice of broadening
scheme.}}
  \label{fig:boke_Appendix}
\end{figure}

As a result, even when only the uncertainty in the imaginary direction is taken into account, the main conclusions of this work remain essentially unchanged. Results of an analysis analogous to that in Fig.~\ref{fig:rhoboke} are shown in Fig.~\ref{fig:boke_Appendix}.

% 参考文献
\bibliography{refs} %ref.bibから拡張子を外した名前

\end{document}